\documentclass{WileyChemistry-template}

\usepackage{amsfonts} 
\usepackage{physics}
\usepackage{nicefrac}       % compact symbols for 1/2, etc.
\usepackage{microtype}      % microtypography
\usepackage{lipsum}
\usepackage{tipx}
\usepackage{bm}% bold math
\usepackage{mhchem}
\usepackage{xr-hyper} % hyperlinks

\makeatletter
\newcommand*{\addFileDependency}[1]{% argument=file name and extension
  \typeout{(#1)}
  \@addtofilelist{#1}
  \IfFileExists{#1}{}{\typeout{No file #1.}}
}
\makeatother

\makeatletter
\newcommand*{\addAuxFileDependency}[1]{% argument=file name and extension
  \makeatletter\@input{x#1.tex}\makeatother
}
\makeatother

\newcommand*{\myexternaldocument}[1]{%
    \externaldocument[#1:]{#1}%
    \addAuxFileDependency{#1}% hu 
    \addFileDependency{#1.tex}%
    \addFileDependency{#1.aux}%
}

\title{\raggedright Self-Consistent Convolutional Density Functional Approximations: Application to Adsorption at Metal Surfaces}

\author{
\begin{minipage}{\textwidth}
	Sushree Jagriti Sahoo\textsuperscript{a}, Qimen Xu\textsuperscript{b}, Xiangyun Lei\textsuperscript{c}, Daniel Staros\textsuperscript{d}, Gopal R. Iyer\textsuperscript{d}, Brenda Rubenstein\textsuperscript{d}, Phanish Suryanarayana\textsuperscript{e}, Andrew J. Medford*\textsuperscript{a}
\end{minipage}
}

\newcommand{\affiliation}{
\begin{itemize}

\item[{a}] Sushree Jagriti Sahoo,  Dr. Phanish Suryanarayana, Dr. Andrew J. Medford*\\
Georgia Institute of Technology, Atlanta, GA\\
Corresponding author's e-mail: ajm@gatech.edu

\item[{b}] Dr. Qimen Xu\\
Georgia Institute of Technology, Atlanta, GA \\
National Supercomputing Center, Shenzhen, People's Republic of China

\item[{c}] Dr. Xiangyun Lei\\
Toyota Research Institute, Los Altos, CA

\item[{d}] Daniel Staros, Gopal R. Iyer, Dr. Brenda Rubenstein\\
Department of Chemistry, Brown University, Providence, RI
\end{itemize}
}

\newcommand{\edens}{\ensuremath{\rho(\mathbf{r})}}
\newcommand{\spinedens}{\ensuremath{\rho_\sigma(\mathbf{r})}}
\newcommand{\edensup}{\ensuremath{\rho_{\uparrow}(\mathbf{r})}}
\newcommand{\edensdown}{\ensuremath{\rho_{\downarrow}(\mathbf{r})}}
\newcommand{\edensspin}[1][\sigma]{\ensuremath{\rho_{#1}(\mathbf{r})}}
\newcommand{\stencils}[1][i,j]{\ensuremath{Z_{#1}}}
\newcommand{\convs}[1][i,j]{\ensuremath{\zeta_{#1}}}
\newcommand{\dpa}{\partial}
\newcommand{\qp}{\textrm{\textqplig}}

\newcommand{\anaint}[2][\mathbf{r}]{\ensuremath{\int #2 \mathrm{d}#1}}
\newcommand{\anaconv}[4][\mathbf{q}]{\ensuremath{\int} #2 (#1) #3 (#4 - #1) \mathrm{d}#1}

\newcommand{\anaintconv}[2][\mathbf{r}]{\ensuremath{\int} #2 \mathrm{d}#1}

\newcommand{\conv}[3][\mathbf{r}]{\ensuremath{\left(#2 \circledast #3 \right)(#1)}}
\newcommand{\shortconv}[3][\mathbf{r}]{\ensuremath{\left(#2 \circledast #3 \right)}}

\newcommand{\epstozero}[1]{\left. #1 \right|_{\varepsilon = 0}}

\renewcommand{\abstract}{The exchange-correlation (XC) functional in density functional theory is used to approximate multi-electron interactions. A plethora of different functionals are available, but nearly all are based on the hierarchy of inputs commonly referred to as ``Jacob's ladder.'' This paper introduces an approach to construct XC functionals with inputs from convolutions of arbitrary kernels with the electron density, providing a route to move beyond Jacob's ladder. We derive the variational derivative of these functionals, showing consistency with the generalized gradient approximation (GGA), and provide equations for variational derivatives based on multipole features from convolutional kernels. A proof-of-concept functional, PBEq, which generalizes the PBE$\alpha$ framework where $\alpha$ is a spatially-resolved function of the monopole of the electron density, is presented and implemented. It allows a single functional to use different GGAs at different spatial points in a system, while obeying PBE constraints. Analysis of the results underlines the importance of error cancellation and the XC potential in data-driven functional design. After testing on small molecules, bulk metals, and surface catalysts, the results indicate that this approach is a promising route to simultaneously optimize multiple properties of interest.
}

\newcommand{\keywords}{
	Density Functional Theory \textbullet\ 
	Exchange-correlation  \textbullet\ 
	Generalized gradient approximation \textbullet\ 
    Multipole expansion
}

\myexternaldocument{supp}
\begin{document}
%%%%%%%%%%%%%%%%%%%%%%%%%%%%%%%%%%%%%%%%%%%%%%%%%%%%%%%%%%
%%%%%%%%%%%%%%%%%%%%%%%%%%%%%%%%%%%%%%%%%%%%%%%%%%%%%%%%%%
%%%%%%%%%%%%%%%%%%%%%%%%%%%%%%%%%%%%%%%%%%%%%%%%%%%%%%%%%%

\twocolumn[\vspace{-1.5cm}\maketitle\vspace{-1cm}
	\textit{\dedication}\vspace{0.4cm}]
\small{\begin{shaded}
		\noindent\abstract
	\end{shaded}
}

\begin{figure} [!b]
\begin{minipage}[t]{\columnwidth}{\rule{\columnwidth}{1pt}\footnotesize{\textsf{\affiliation}}}\end{minipage}
\end{figure}

%%%%%%%%%%%%%%%%%%%%%%%%%%%%%%%%%%%%%%%%%%%%%%%%%%%%%%%%%%
%%%%%%%%%%%%%%%%%%%%%%%%%%%%%%%%%%%%%%%%%%%%%%%%%%%%%%%%%%
%%%%%%%%%%%%%%%%%%%%%%%%%%%%%%%%%%%%%%%%%%%%%%%%%%%%%%%%%%

%%%%%%%		 Main Text			%%%%%%% 

%	For Communications for Angewandte Chemie, please remove headlines for Introduction, Results and Discussion and Conclusion

\section{Introduction}
\label{introduction}

\label{sec:Introduction}

Density functional theory (DFT) is a widely used computational tool in the fields of computational chemistry, materials science, and solid-state physics that aims to approximate the Schr\"{o}dinger equation in order to predict the electronic structure and properties of chemical systems.~\cite{Hohenberg1964InhomogeneousGas,Kohn1965Self-ConsistentEffects} It is a powerful tool for investigating the electronic and structural properties of materials due to its excellent balance between speed and accuracy. DFT is based on the Hohenberg-Kohn theorem,~\cite{PhysRev.136.B864} which states that the ground state energy of a system is a unique functional of the electron density. It utilizes a set of single-particle equations known as the Kohn-Sham (KS) equations that are solved self-consistently to determine the total energy and other properties of the system. The exchange-correlation (XC) functional, $E_{xc}$, in DFT is the main approximation that accounts for the multi-electron interactions in a system, but its universal form is not known, making it necessary to select different approximations for different types of systems. Selecting the most appropriate XC functional for a given application is one of the main challenges in DFT.\cite{Medvedev2017DensityFunctional,teng2014choosing,rappoport2006approximate,ostrom2022designing}

The formulation of XC functionals requires the construction of a model space to accurately represent the electronic environment of a chemical system, and a functional to map the electronic environment to the energy density. Different classes of functionals with increasing complexity of model space are captured by the popular analogy of ``Jacob's ladder''.~\cite{Perdew2001JacobsEnergy} The first rung of the ladder is occupied by the local density approximation (LDA) where Kohn and Sham approximated $E_{xc}$ as a function of the local (spin) electron density of a uniform electron gas. This approximation is known to be relatively accurate for delocalized systems such as metals.~\cite{Kohn1965Self-ConsistentEffects} The second rung is occupied by the generalized gradient approximation (GGA) which includes information about the gradient of the electron density.~\cite{Langreth1981EasilyFunctional,Becke1988CorrelationModel,Perdew1986AccurateApproximation} Different ways of including this information result in a large number of distinct GGA functionals that include the widely used Perdew-Burke-Ernzerhof (PBE)~\cite{Perdew1996GeneralizedSimple} and Perdew-Wang (PW91) functionals.~\cite{Sholl2009DensityIntroduction} The next improvement over GGA functionals is the class of meta-GGA (mGGA) functionals that include the kinetic energy density.~\cite{metaGGA} These functionals include a range of empirical and physical approximations, but improvements are not always systematic.~\cite{Medvedev2017DensityFunctional} The mGGA family includes the popular Strongly Constrained and Appropriately Normed (SCAN) semilocal density functional and its revisions.~\cite{Sun2015StronglyFunctional,rSCAN,r2SCAN} The next class of functionals are the ``hybrid functionals'' that include a component of fully non-local ``exact'' exchange based on the Fock operator.~This class includes functionals such as B3LYP,~\cite{Becke1993Density-functionalExchange,Lee1988DevelopmentDensity} PBE0,~\cite{Adamo1999TowardModel} and HSE06.~\cite{Heyd2003HybridPotential,Krukau2006InfluenceFunctionals} Each ``rung'' of Jacob's ladder promises to improve the general accuracy of DFT calculations by including more exact constraints and non-local information from orbital densities.~\cite{Peverati2011Communication:Chemistry,Perdew2008DensitySatisfaction,Precechtelova2014Communication:Energy} However, the accuracy of XC functionals does not improve systematically for all systems, making it difficult to establish an \textit{a priori} estimate of the accuracy of DFT.~\cite{cohen2012challenges,dftchallengereview,zhao2008density,kalita2021learning, Medvedev2017DensityFunctional}

Due to the lack of a universal functional, a variety of XC functionals have been developed and evaluated for different applications.~\cite{Kohn1965Self-ConsistentEffects,Langreth1981EasilyFunctional,Becke1988CorrelationModel,Perdew1986AccurateApproximation,Sun2015StronglyFunctional,Becke1993Density-functionalExchange,Lee1988DevelopmentDensity,Adamo1999TowardModel,Heyd2003HybridPotential,Krukau2006InfluenceFunctionals} However, the limited transferability of these functionals across different systems poses a significant challenge. In many cases, chemical systems have diverse electronic environments that cannot be effectively captured by a single XC functional. For example, adsorption of a gas molecule on a solid surface requires accurate treatment of bulk, surface, and molecular electronic environments and may involve metallic, ionic, covalent, hydrogen, and van der Waals bonding simultaneously.\cite{BEEF-vdW} Nonetheless, DFT is commonly used to study surface adsorption systems because of its ability to treat large and (semi)periodic systems that are impractical with highly accurate wavefunction theories (WFTs),~\cite{zhang2019coupled} and the experimental difficulties associated with measuring adsorption energies. Several functionals have been developed specifically for surface science applications, including the RPBE functional,~\cite{RPBE} the Bayesian error estimation (BEEF) family of functionals~\cite{BEEF-vdW,mBEEF}, and the Armiento and Mattsson (AM05) functional.~\cite{Armiento2005FunctionalTheory} However, even with these tailored XC functionals, it has been shown that adsorption energies may vary by up to 1 eV in some cases, which can lead to qualitatively incorrect conclusions in the analysis of surface reactions.\cite{kepp2018accuracy,WELLENDORFF201536}

Several approaches have been used to improve the performance of functionals for surface systems. In the case of RPBE, Hammer and N\o{}rskov elegantly showed that a small modification to the functional ansatz of PBE improves the performance for adsorption systems while still obeying the physical constraints of PBE.~\cite{RPBE} Another approach, introduced by Kohn and co-workers, utilizes the concept of subsystem functionals with the edge electron gas model. This model explains the shortcomings of existing approaches like LDA and GGA for regions where the nature of KS wave functions transitions from propagating to evanescent.~\cite{Kohn1998EdgeGas} %These functionals perform well for bulk systems with near-homogeneous density. 
Several strategies have been proposed to formulate functionals based on this model, such as correcting the surface energies of LDA and GGA by describing the effective potential as an exponential,~\cite{Mattsson2001AnSurfaces} the Mathieu gas model that uses the effective cosine potential,~\cite{Armiento2002SubsystemParticle} and the AM05 functional designed to include surface effects.~\cite{Armiento2005FunctionalTheory} The underlying idea is to divide a system into sub-systems and employ different functionals to approximate the electronic structure of each part accurately. More recently, the BEEF family of functionals took a data-driven approach, fitting the best available data across a range of molecular, bulk, and surface properties.~\cite{BEEF-vdW,mBEEF} The resulting functionals utilize statistical techniques to provide an error estimate based on perturbations to the fitted parameters. The design of these functionals also quantified a Pareto optimal between different properties, indicating that for a given model space, it is not possible to simultaneously improve performance for all metrics of interest.

In recent years, the data-driven paradigm has been increasingly applied to the problem of XC functional design. Several neural network and deep learning models have emerged, leveraging inputs from the model space of Jacob's ladder, showing promise in the development of these functionals. These models make use of inputs such as the electron density, density gradients, kinetic energy density, Fock energy, and more. Notable contributions in this area include the work by Burke and colleagues,\cite{snyder2012finding,snyder2013orbital,li2016understanding} in which machine learning (ML) is used to estimate the kinetic energy of non-interacting electrons, bypassing the need to solve the KS equations. Recently, DeepMind published their state-of-the-art deep learning models that adhere to specific constraints related to fractional electrons.~\cite{DM21} Other researchers have used symbolic functional evolutionary searches, physically constrained neural networks, and data augmentation to satisfy known physical limits.\cite{SyFES_functional,physicalconstraint,Gedeon_2022,D3DD00114H} Another notable work in this field is by Nagai et al., which uses a feed-forward neural network that takes inputs from LDA, GGA, and meta-GGA model spaces, along with a new ``near-region approximation'' (NRA) input to construct an XC functional trained on a few reference molecules that is applicable to hundreds of molecules.~\cite{nagai2020completing} In another study, Marivi Fernandez-Serra and colleagues introduced a novel framework called NeuralXC which leverages supervised ML to construct density functionals that explicitly depend on the electron density and implicitly incorporate atomic positions.~\cite{dick2020machine} The authors adopt a $\Delta$-learning type approach and demonstrate how variational derivatives can be computed, thereby creating ML density functionals that can be used in self-consistent calculations. This functional significantly improves the accuracy beyond the GGA level, approaching that of coupled cluster theory, and is transferable across diverse systems but relies on the use of localized atomic orbitals and has not been demonstrated on periodic systems. These early successes indicate the promise of ML approaches for XC design, but the reliability and transferability of ML-based functionals are influenced by the quality and representations used for the training dataset. Therefore, careful selection of training data and incorporation of physical principles into the ML framework are essential to ensure the robustness of these functionals.

The vast majority of work on XC functional design has been limited to inputs or model spaces defined by Jacob's ladder. The fields of computer vision and ML suggest an alternative paradigm, where 3D convolutional kernels are used to describe the electron density, which is effectively a 3D image projected onto a finite-difference grid. Several studies have explored convolutional descriptors and neural networks for XC functionals.~\cite{Lei2019DesignDescriptors,gong2022incorporation} However, the lack of variational derivatives is a practical challenge that prevents these functionals from being used in self-consistent calculations in DFT, particularly for periodic and semi-periodic systems. Recently, several studies have explored fully self-consistent ML functionals that go beyond Jacob's ladder. Kresse and colleagues recently used a power spectrum representation of the electron density along with Gaussian process regression to estimate the exchange enhancement factor corresponding to reference data from the random-phase approximation. \textcolor{black}{This functional is implemented in planewave DFT and necessitates the use of Fourier transforms for calculation of variational derivatives.~\cite{riemelmoser2023machine}} In addition, Bystrom \& Kozinsky have introduced two scale-invariant nonlocal descriptors of electron density and used them to train a ML model that satisfies the uniform scaling rule for exchange energy. \textcolor{black}{These works have an implementations for a planewave basis and Gaussian-type orbitals.}~\cite{bystrom2022cider,bystrom2023nonlocal}  These studies demonstrate the powerful potential of novel electron density descriptors, \textcolor{black}{but at present the calculations for periodic systems are limited to planewave DFT.} %codes due to the Fourier transforms used in the variational derivative.

In this study, we first derive a general variational derivative for functionals constructed from arbitrary convolutional kernels, with a result that does not rely on any specific basis set or Fourier transforms. We then utilize the systematically expandable model space introduced by Lei et al.,~\cite{Lei2019DesignDescriptors} which is constructed by extracting fingerprints using 3D convolutional kernels at varying length scales, to derive and implement a proof-of-concept functional in the state-of-the-art SPARC DFT code.\cite{ghosh2017sparc1, Xu2021Sparc:Calculations,  zhang2023version} \textcolor{black}{We note that the current implementation is only applicable to periodic cells, but can be easily extended to non-periodic systems within the same framework.}
%We note that the convolutional descriptors are not scale-invariant, therefore the functional does not satisfy the uniform density scaling constraint for $E_x$. 
The functional is constructed by generalizing the formulation of the PBE$\alpha$ functional such that $\alpha$ is a function of an arbitrary electron density descriptor.\textcolor{black}{\cite{PhysRevB.75.195108}} This enables the use of different approximations at different points in space within a given system to distinguish between different electronic environments present in a single system. We evaluate the capabilities of this functional to calculate the properties of small organic molecules and bulk metals. We also test this functional for small chemisorption systems to assess its ability to calculate adsorption energies in systems with an interface between metallic and molecular systems by evaluating binding site preference of CO on (111) surfaces of Pt, Cu, and Rh. We include a detailed comparison with Quantum Monte Carlo (QMC) results based on Diffusion Monte Carlo (DMC) calculations for the Pt(111) system. 
DMC provides highly accurate adsorption energies in chemisorbed systems.\cite{iyer_rubenstein_dmc, doblhoff_dier_dmc, powell_dmc} This affords us energetic resolution of distinct bonding environments at a level that is inaccessible to purely experimental studies.\cite{iyer_rubenstein_dmc, hsing_dmc}
The results reveal that our functional is capable of correctly distinguishing the correct adsorption site of CO on Pt(111), and is at least as accurate as RPBE and other GGA functionals for the systems studied. Overall, the results of the work provide both general and specific insights into the design of XC functionals for chemistry, materials, and surface science.
%Comparisons with DMC adsorption energies, therefore, provide a pathway to extending the accuracy of the generalized functional developed herein to the level of QMC methods and other many-body electronic structure theories at significantly lower computational expense. Here, we include DMC adsorption energies for comparison wherever applicable.}

\section{Results and Discussion}
\label{sec:Results}

\subsection{Theoretical formulation}
\label{subsec:Theory}

\subsubsection{Variational derivative of convolutional functionals}

The variational derivative (also known as the functional derivative or Gateaux derivative) of a given functional $E[\rho]$ is given by $\frac{\delta E}{\delta \rho}$:~\cite{zbMATH00837932}
\begin{equation}
\label{eq:definition}
\begin{aligned}
    \int \frac{\delta E}{\delta \rho}(\mathbf{r})y(\mathbf{r}) \mathrm{d}\mathbf{r} &= \left.\frac{\dpa}{\dpa\epsilon}E[\rho + \epsilon y]\right|_{\epsilon = 0}\,,
\end{aligned}
\end{equation}
where $y$ is an arbitrary function and $\varepsilon y$ is the variation with respect to the electron density, $\rho$. Consider a generic energy functional given by $E[\rho]$, where:
\begin{equation}
    \label{eq:convFunctional}
        {E[\rho] = \anaint{f(\boldsymbol{\lambda}[\rho](\mathbf{r}))}}\,,
\end{equation}
where $\edens$ is the spin-paired electron density, $f$ is an arbitrary function, and $\boldsymbol{\lambda} = \{\lambda_0,\lambda_1..., \lambda_n\}$ are ``descriptors'' computed from the electron density. We note that vector quantities are represented by bold symbols and their components by indices, e.g., $\boldsymbol{\lambda}$ or $\lambda_i$, respectively.  In this work, we consider descriptors calculated from convolutions of $\edens$ and arbitrary convolutional kernels:
\begin{equation} \label{eq:zeta:def}
    \begin{aligned}
       \zeta_{i,j}[\rho](\mathbf{r})  = \conv{\stencils}{\rho} = &\anaconv{\stencils}{\rho}{\mathbf{r}}\\
        = & \anaconv{\rho}{\stencils}{\mathbf{r}}\,,
    \end{aligned}
\end{equation}
where \stencils{} corresponds to the $j^{\mathrm{th}}$ 3-dimensional convolutional kernel used to construct the $i^{\mathrm{th}}$ descriptor, \convs{} represents the result of the convolution between the kernel and the electron density. The domain of the convolutional integral in this case must only span the domain of the associated kernel, \stencils{}, and will thus be less computationally expensive than an integral over all space (assuming the kernels are semilocal). For generality, we consider descriptors computed from transformations of multiple convolutional outputs, where the $i^{\mathrm{th}}$ descriptor can be computed as:
\begin{equation}
\label{eq:descriptor}
        \lambda_i = g_i(\convs[i,0], \convs[i,1] , ... \convs[i,M]) \,,
\end{equation}
where $g_i$ is an arbitrary function corresponding to the  $i^{\mathrm{th}}$ descriptor. We note that this intermediate function is introduced because it is common to need non-linear transforms (e.g. $L_2$-norm) of individual convolutions to ensure that descriptors are rotationally invariant. %Finally, we can write the generic form of the functional as:

%\begin{equation}
%E[\rho] = \anaint{f(\lambda_i(\convs(\edens))} = \anaint{f(\lambda_i(\mathbf{r})}\,.
%\end{equation}

Applying the definition of the variational derivative (Eq.~\ref{eq:definition}) to the generic form of the convolutional functional (Eq.~\ref{eq:convFunctional}) yields:
\begin{equation}
\label{eq:vardiffdef}
    \begin{split}
        \epstozero{\frac{\dpa }{\dpa \varepsilon} E[\rho + \varepsilon y]} = 
        \epstozero{\frac{\dpa }{\dpa \varepsilon} \anaint{f(\boldsymbol{\lambda}([\rho+ \varepsilon y](\mathbf{r})
        )}}\\
        = 
        \epstozero{\anaint{\sum_i \frac{\dpa f}{\dpa \lambda_i}(\mathbf{r}) 
        \sum_j \frac{\dpa \lambda_i}{\dpa \zeta_{i,j}}(\mathbf{r})
        \frac{\dpa
        \zeta_{i,j}[\rho + \varepsilon y](\mathbf{r}) }{\dpa \varepsilon} 
        }}
        \,.
    \end{split}
\end{equation}
It is straightforward from the definition of $\zeta_{i,j}$ (Eq.~\ref{eq:zeta:def}) that it is a linear functional, i.e., $\zeta_{i,j}[c_1 f_1 + c_2 f_2] = c_1 \zeta_{i,j}[f_1] + c_2 \zeta_{i,j}[f_2]$. Thus we have:
% \begin{equation}
%     \zeta_{i,j}[c_1 f_1 + c_2 f_2] = c_1 \zeta_{i,j}[f_1] + c_2 \zeta_{i,j}[f_2] \,.
% \end{equation}
\begin{equation}
\begin{aligned}
    \frac{\dpa \zeta_{i,j} [\rho + \varepsilon y](\mathbf{r})}{\dpa \varepsilon} 
    &= \zeta_{i,j}[y](\mathbf{r})\,.
\end{aligned}
\end{equation}
Substituting into Eq.~\ref{eq:vardiffdef} yields:
\begin{equation}
    \label{eq:func_deriv_form}
    \begin{split}
    \epstozero{\frac{\dpa }{\dpa \varepsilon} E[\rho + \varepsilon y]} 
    = \epstozero{\anaint{\sum_i \frac{\dpa f}{\dpa \lambda_i}( \mathbf{r} ) 
        \sum_j \frac{\dpa \lambda_i}{\dpa \zeta_{i,j}}(\mathbf{r})
        \zeta_{i,j}[y](\mathbf{r})
    }}  \\
    =  \anaint{\sum_i \frac{\dpa f}{\dpa \lambda_i}( \mathbf{r} ) 
        \sum_j \frac{\dpa \lambda_i}{\dpa \zeta_{i,j}}(\mathbf{r})
        \anaconv{y}{\stencils}{\mathbf{r}}
    }   \\
    = \int \anaintconv[\mathbf{q}]{ \sum_i \frac{\dpa f}{\dpa \lambda_i}(\mathbf{q}) 
        \sum_j \frac{\dpa \lambda_i}{\dpa \zeta_{i,j}}(\mathbf{q})
        Z_{i,j}(\mathbf{q} - \mathbf{r}) }  \, y(\mathbf{r}) \mathrm{d}\mathbf{r} \,. 
        \end{split}
\end{equation}
where we have made use of the fact that $\mathbf{r}$ and $\mathbf{q}$ are dummy variables and can be swapped. Comparing with Eq.~\ref{eq:definition} reveals the variational derivative (or, equivalently, the potential $V(\mathbf{r})$): %In arriving at the last equation, we swapped the dummy variables $\mathbf{r}$ and $\mathbf{q}$. 
\begin{equation}
    \label{eq:funcdiff3}
    \frac{\delta E}{\delta \rho}(\mathbf{r}) \coloneqq V(\mathbf{r}) =  \anaintconv[\mathbf{q}]{\sum_i \frac{\dpa f}{\dpa \lambda_{i}}(\mathbf{q})  \sum_j \frac{\dpa \lambda_{i}}{\dpa \convs[i,j]}(\mathbf{q}) Z_{i,j}(\mathbf{q} - \mathbf{r})}\,,
\end{equation}
Defining $\nu_{i,j}(\mathbf{q}) \equiv \frac{\dpa f}{\dpa \convs[i,j]}(\mathbf{q})$, $Z'_{i,j}(\mathbf{r}) \equiv Z_{i,j}(-\mathbf{r})$, rearranging the summations, and grouping the partial derivatives, Eq.~\ref{eq:funcdiff3} can be expressed as:
\begin{equation}
    \label{eq:funcdiff:rewritten}
    V(\mathbf{r}) = \sum_i \sum_j \anaintconv[\mathbf{q}]{\nu_{i,j}(\mathbf{q}) Z'_{i,j}(\mathbf{r} - \mathbf{q})} \,.
\end{equation}
The integral in Eq.~\ref{eq:funcdiff:rewritten} follows the definition of a convolution which yields the following expression for $V(\mathbf{r})$:
\begin{equation}
\begin{aligned}
    \label{eq:funcdiff4}
    % V(\mathbf{r}) &= \sum_i \sum_j \conv{\frac{\dpa f}{\dpa \convs[i,j]}}{Z'_{i,j}} \\ 
    V(\mathbf{r}) &= \sum_i \sum_j \conv{\nu_{i,j}}{Z'_{i,j}}
    &= \sum_i \sum_j \conv{Z'_{i,j}}{\nu_{i,j}}\,,
\end{aligned}
\end{equation}
where the last equality reflects the fact that the convolution operator commutes. This form reveals that the variational derivative of a convolutional kernel can be decomposed into a term that only requires standard derivatives ($\nu_{i,j}$), but requires an additional number of convolutions equal to the total number of convolutional kernels used to calculate all the descriptors.

The prior derivation is applicable to spin-paired functionals where a single electron density, $\edens$ is present. However, for spin-polarized functionals there exist two independent electron densities for spin-up and spin-down electrons, denoted as $\edensup$ and $\edensdown$, respectively, or as $\edensspin$ where $\sigma$ is a spin index $\sigma \in [\uparrow, \downarrow]$.
In this case, each convolutional feature $\convs$ and corresponding descriptor $\lambda_i$ will have an additional index denoting the spin channel, resulting in $\convs[i,j,\sigma]$ and $\lambda_{i,\sigma}$. In addition, there will be two resulting potentials, $V_{\sigma}$, which will be coupled together, since $f$ will generally be a function of both spin-up and spin-down descriptors:
\begin{equation}
    \begin{aligned}
        E[\rho] &= \anaint{f(\boldsymbol{\lambda}_{\uparrow}[\rho_{\uparrow}](\mathbf{r}), \boldsymbol{\lambda}_{\downarrow}[\rho_{\downarrow}](\mathbf{r}))} \,,\\
    \end{aligned}
\end{equation}
following the derivation above to Eq.~\ref{eq:funcdiff4} yields:
\textbf{\begin{equation}
    \begin{aligned}
    \label{eq:v_potential}
        V_{\sigma}(\mathbf{r}) &= \sum_i \sum_j \conv{\frac{\dpa f}{\dpa \convs[i,j,\sigma]}}{Z'_{i,j}}\\
                            &= \sum_i \sum_j \conv{Z'_{i,j}}{\nu_{i,j,\sigma}}\,,
    \end{aligned}
\end{equation}}
or, explicitly:
\begin{equation}
    \begin{aligned}
    \label{eq:v_up_down}
        V_{\uparrow}(\mathbf{r})  
         &= \sum_i \sum_j \conv{Z'_{i,j}}{\nu_{i,j,\uparrow}}\,, \\
          V_{\downarrow}(\mathbf{r})  
          &= \sum_i \sum_j \conv{Z'_{i,j}}{\nu_{i,j,\downarrow}}\,. 
    \end{aligned}
\end{equation}
We note that the standard GGA can be expressed as a special convolutional functional if the gradient is computed using finite-difference stencils, and show that the standard expression for the GGA potential is recovered in this case (see SI). 

\subsubsection{Potential for a functional based on multipole descriptors}
\label{sec:funcderivMP}

Multipole (MP) descriptors are a class of rotationally invariant descriptors based on the concept of multipole expansions of the electron density that has been ``screened'' by radial probes \cite{Lei2019DesignDescriptors, Lei2022}. Different radial probes such as heaviside step functions or Legendre polynomials can be utilized. The MP descriptors developed thus far follow the form:

\begin{equation}
    \qp_{l,m,\sigma}(\mathbf{r}) = \sqrt{\sum_j \omega_j \cdot \left( \conv{R_{l}M_{m,j}}{\rho_\sigma} \right)^2}\,,
\end{equation}
where $l$ indexes the radial probe, $m$ refers to the order of the multipole, $\omega_j$ is a weight for each term, $R_l$ is the radial probe kernel, $M_{m,j}$ is a kernel based on spherical harmonics, with $j$ indexing the different possible groups of a given rotational order $m$. For simplicity, the radial and angular indices $l$ and $m$ can be combined into an index $i$ that covers all combinations of $l$ and $m$. Thus, the radial and angular kernels $R_l$ and $M_{m,j}$ can also be combined into a set of kernels, \stencils, where $i$ refers to the $i^{\mathrm{th}}$ MP feature and $j$ corresponds to a set of kernels that arise from the different rotational symmetry groups and must be normed. Thus, the MP descriptors can be written as a specific case of Eq.~\ref{eq:descriptor}:
\begin{equation}
    \begin{aligned}
    \lambda_{i,\sigma}(\mathbf{r}) \coloneqq \qp_{i,\sigma}(\mathbf{r}) =& \sqrt{\sum_j \omega_j \cdot \left( \conv{\stencils}{\rho_\sigma} \right)^2} \\
    =& \sqrt{\sum_j \omega_j \zeta_{i,j,\sigma}^2}\,.
    \end{aligned}
\end{equation}
This form is general to any feature constructed by taking the $L_2$-norm of a vector obtained from a linear combination of multiple convolutional kernels. The partial derivative of $\qp_{i,\sigma}$ with respect to $\convs[i,k,\sigma]$ is:
\begin{equation}
    \begin{aligned}
    \frac{\dpa \qp_{i,\sigma}}{\dpa \convs[i,k,\sigma]} &= \frac{\frac{\dpa}{\dpa \convs[i,k,\sigma]} \sum_j \omega_j \convs[i,j,\sigma]^2 }{2 \sqrt{\sum_j \omega_j \convs[i,j,\sigma]^2}} \\
    &= \frac{ \sum_j \omega_j \convs[i,j,\sigma] \delta_{j,k} }{\sqrt{\sum_j \omega_j \convs[i,j,\sigma]^2}}\\
    &= \frac{ \omega_k \convs[i,k,\sigma] }{\qp_{i,\sigma}}\,,
    \end{aligned}
\end{equation}
substituting into Eq.~\ref{eq:funcdiff3} and simplifying yields:
\begin{equation}
    \label{eq:MPdiff}
    V_\sigma(\mathbf{r}) = \sum_i \sum_j \omega_j \conv{Z'_{i,j}} {\frac{\dpa f}{\dpa \qp_{i,\sigma}} \frac{\convs[i,j,\sigma]}{\qp_{i,\sigma}}}\,,
\end{equation}
comparing to Eq.~\ref{eq:funcdiff4} reveals that for the case of MP features:
\begin{equation}
    \label{eq:MPnu}
    \nu_{i,j,\sigma}(\mathbf{r}) = \omega_j \frac{\dpa f(\mathbf{r})}{\dpa \qp_{i,\sigma}} \frac{\convs[i,j,\sigma](\mathbf{r})}{\qp_{i,\sigma}}\,,
\end{equation}
where the intermediate quantities required, $\qp_{i,\sigma}$ and $\zeta_{i,j,\sigma}$, must already be calculated in the calculation of the descriptors, and the derivative $\frac{\dpa f}{\dpa \qp_{i,\sigma}}$ can be calculated analytically or by automatic differentiation depending on the functional form and implementation of $f$.

\subsubsection{PBEq: A class of physically constrained functionals based on a convolutional input}
% \subsubsection{Functional form of $\alpha(\mathbf{r})$ in exchange enhancement factor of PBE$\alpha(\qp)$}

To implement and numerically evaluate the convolutional potential derivation, it is necessary to define a specific form of the descriptors and the functional. In this work, we provide a proof-of-concept by extending the formulation of the exchange enhancement factor of the functional form of PBE$\alpha$ \cite{PhysRevB.75.195108} so that $\alpha$ is a function of a single MP descriptor denoted by $q$. The original PBE$\alpha$ functional is a generalized form of the widely used PBE functional~\cite{Perdew1996GeneralizedSimpleb} that can be systematically varied by a single adjustable parameter, $\alpha$, which is a constant in the original formalism. The PBE$\alpha$ functional obeys all exact constraints of PBE for all values of $\alpha$, and the simple form of PBE$\alpha$ leads to a smooth exchange potential. PBE$\alpha$ is equivalent to PBE and RPBE when $\alpha = 1$ and $\alpha \to \infty$,  respectively \cite{PhysRevB.75.195108}. When $\alpha = 0.52$, the functional obeys the second-order expansion of the slowly-varying limit derived by Wu and Cohen (WC) ~\cite{PhysRevB.73.235116}, and performs similarly to the WC functional for densely packed solids \cite{PhysRevB.75.195108}. 

We begin with a brief review of the formulation of the PBE$\alpha$ functional. The PBE$\alpha$ functional and our modifications to it provide only the exchange energy density, and the correlation energy is identical to PBE.  The general form of the exchange energy density per particle for a GGA functional is the product of $\spinedens$, $\varepsilon_{x}^{LDA}(\spinedens)$ and the exchange enhancement factor, represented by $F_x(s_\sigma(\mathbf{r}))$, where $s_\sigma(\mathbf{r})$ is the ``reduced'' gradient which is given by:

\begin{equation}
    \label{eq:s_sgima}
    s_{\sigma}(\mathbf{r}) = \frac{| \grad \rho_{\sigma}(\mathbf{r})|}{2 (6\pi^2)^{1/3} \rho_{\sigma}(\mathbf{r})^{4/3}}\,,
\end{equation}
where $| \grad \rho_{\sigma}(\mathbf{r})|$ is the $L_2$-norm of the gradient vector: \\$| \grad \rho_{\sigma}(\mathbf{r})| = \sqrt{\sum_{j=1}^3 (\nabla_j \rho_{\sigma}(\mathbf{r}))^2}$. \newpage For simplicity, we represent $\varepsilon_{x}^{GGA}(\spinedens,s_\sigma(\mathbf{r}))$ as:
\begin{equation}
    \varepsilon_{x,\sigma}^{GGA} = \rho_\sigma\varepsilon_{x,\sigma} F_{x,\sigma}\,,
\end{equation}
where $\varepsilon_{x,\sigma}$ is the LDA exchange energy density per particle. $F_{x,\sigma}$ in PBE$\alpha$ is given by:

\begin{equation}
    \begin{aligned}
        F_{x,\sigma}^{PBE_\alpha} = 1 + \kappa\left[1 - \frac{1}{\left(1+ \frac{\mu s_\sigma^2}{\kappa\alpha}\right)^\alpha}\right]\,.
    \end{aligned}
\end{equation}
The spin-dependent exchange potential for PBE$\alpha$, $V_{x,\sigma}^{PBE\alpha}$ is given by:

\begin{equation}
\begin{aligned}
    \label{eq:convPBEalpha}
     V_{x,\sigma}^{PBE\alpha} = \frac{\dpa\varepsilon_{x,\sigma}^{PBE\alpha}}{\dpa \rho_\sigma} - \sum_{j=1}^{3}{\left[\nabla_j\left(\frac{\dpa  \varepsilon_{x,\sigma}^{PBE\alpha}}{\dpa\nabla_j \rho_\sigma}\right)\right]}\,,
\end{aligned}
\end{equation}
where $\nabla_j$ is the gradient operator that can be represented by a finite-difference stencil and its convolutional kernel in the case of real-space DFT. The terms $\frac{\dpa  \varepsilon_{x,\sigma}^{PBE\alpha}}{\dpa\nabla_j \rho_\sigma}$ and $\frac{\dpa  \varepsilon_{x,\sigma}^{PBE\alpha}}{\dpa\nabla_j \rho_\sigma}$ can be expanded as the following expressions (see Sec.\ref{supp:sec:pbealpha_derivation} in SI for full derivation):

\begin{equation}
    \label{eq:deriv_rho}
    \frac{\dpa\varepsilon_{x,\sigma}^{PBE\alpha}}{\dpa \rho_\sigma} = \varepsilon_{x,\sigma}\left(\frac{4}{3}F_{x\sigma} +\mu\left(1+\frac{\mu s_\sigma^2}{\kappa\alpha}\right)^{-\alpha - 1}\left(\frac{-8s_\sigma^2}{3}\right) \right)\,,
\end{equation}

\begin{equation}
    \label{eq:deriv_gradrho}
    \frac{\dpa  \varepsilon_{x,\sigma}^{PBE\alpha}}{\dpa\nabla_j \rho_\sigma} = \frac{ {(\nabla_j \rho_\sigma)} \varepsilon_{x\sigma} \mu s_\sigma}{ (6\pi^2)^{1/3} \rho_\sigma^{1/3} |\grad \rho_{\sigma}|} \left(1+\frac{\mu s_\sigma^2}{\kappa\alpha}\right)^{-1 - \alpha}
\end{equation}

\begin{figure}[ht]
\hspace*{-3cm} 
\begin{center}
\includegraphics[width=7cm]{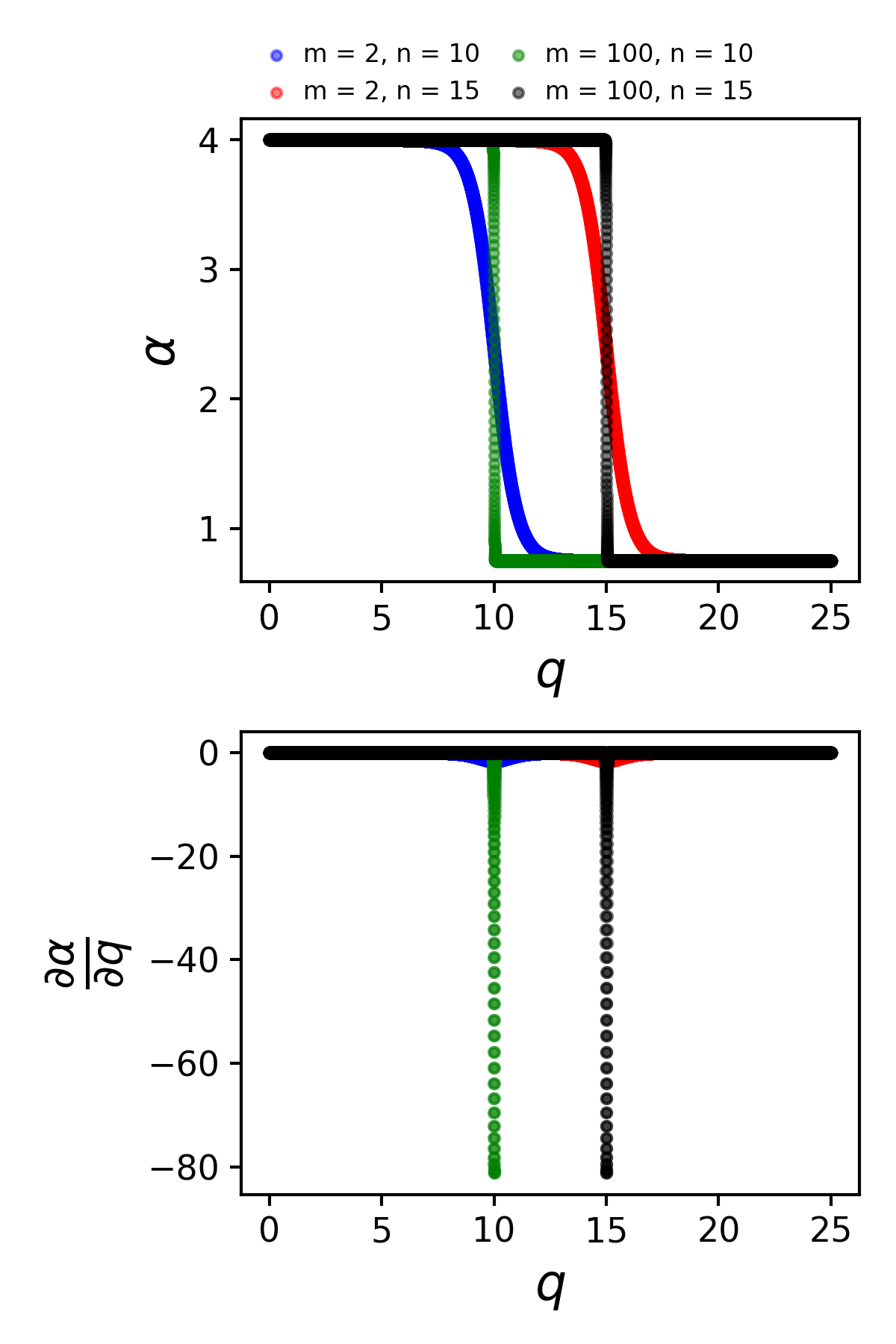}
\vspace*{-5mm}
\caption{Dependence of $\alpha$ on monopole feature, $q$ for different values of m and n (top) and derivative of $\alpha$ w.r.t $q$ (bottom).}
\label{fig:alpha_deriv}
\end{center}
\end{figure}

% \begin{figure*}
%     \centering
%     \includegraphics[keepaspectratio=true,scale=0.6,clip=True]{latex/sigmoid_deriv.png}%wid\thetah=1.\linewid\thetah
%     \caption{Functional form of $\alpha$ with respect to monopole feature, represented by $q(\mathbf{q})$ for different values of m and n}\label{fig:alpha_deriv}
% \end{figure*}

 By constructing $\alpha$ such that it is dependent on MP features, our objective is to demonstrate a functional where $\alpha$ varies for different points in space based on convolutional inputs. 
 %We leverage the potential of MP features in capturing electronic structure information by performing arbitrary convolutions on self-consistent electron density within a DFT calculation. 
For simplicity, we restrict the dependence to a single MP feature, $q_{\sigma}(\mathbf{r})$, such that $\alpha_\sigma$ takes the following form:
\begin{equation}
    \begin{aligned}
        \alpha_\sigma(\mathbf{r}) = g(q_{\sigma}(\mathbf{r}))\,.
    \end{aligned}
\end{equation}
%where $q_{\sigma}(\mathbf{r})$ is a spatially-resolved property dependent on self-consistent electron density from DFT calculations. 
%We can select different options for $q(\mathbf{r})$ based on the angular and radial order of multipole feature. Therefore, 
We denote the corresponding framework as PBEq, where the total exchange energy can be written as a function of $\rho_\sigma$, $\grad \rho_\sigma$, and $q_{\sigma}$:
\begin{equation}
     E_x[\rho]^{PBEq} = \anaint{f(\rho_\sigma, \grad \rho_\sigma, q_{\sigma})}\,.
\end{equation}
The resulting functional will, by construction, satisfy \textcolor{black}{nearly all of the same} exact constraints as PBE, \textcolor{black}{since the constraints are satisfied based on the form of the function for the exchange enhancement factor for all positive values of $\alpha$.} The exception is the density scaling constraint, which will be violated because $q_{\sigma}$ (and the multipole features in general) are not scale invariant. This restriction may be overcome by using alternative features \cite{bystrom2022cider, bystrom2023nonlocal} or more advanced machine learning models \cite{gong2022incorporation}, but these extensions are beyond the scope of this work.

By comparing the general definition of the potential of a convolutional functional (Eq.~\ref{eq:v_potential}) and a GGA functional (Eq.~\ref{eq:convPBEalpha}), it is straightforward to show that:
\begin{equation}
\label{eq:convPBEalphamultipole}
    \begin{aligned}
        V_{x,\sigma}^{PBEq}  = V_{x,\sigma}^{PBE\alpha} 
        + V_{x,\sigma}^{q}\,,
    \end{aligned}
\end{equation}
where
\begin{equation}
    \begin{aligned}
        V_{x,\sigma}^{q} = \shortconv{Z_q}{\frac{\dpa f}{\dpa q_\sigma} }\,,
    \end{aligned}
\end{equation}
and $\frac{\dpa f}{\dpa q_\sigma}$ is given by the following expression:
\begin{equation}
    \begin{aligned}
        \frac{\dpa f}{\dpa q_\sigma} = -\kappa\rho_\sigma \varepsilon_{x,\sigma}\left(1 + \frac{\mu s_\sigma^2}{\kappa\alpha_\sigma}\right)^{-\alpha_\sigma}\\ \left[-\ln\left(1 + \frac{\mu s_\sigma^2}{\kappa\alpha_\sigma}\right) +{\frac{\mu s_\sigma^2}{\kappa\alpha_\sigma}}\left(\frac{1}{1+\frac{\mu s_\sigma^2}{\kappa\alpha_\sigma}}\right)\right]\frac{\dpa\alpha_\sigma}{\dpa q_\sigma}\,.
    \end{aligned}
\end{equation}

\paragraph{}
As a proof-of-concept, we seek to use the revised formulation to develop a functional that switches between two known GGA functionals depending on local environment, where the local environment is represented by a multipole descriptor. This will be achieved using a sigmoid form of $\alpha_\sigma$:
\begin{equation}
\label{eq:alphafunctionalform}
    \begin{aligned}
       \alpha_\sigma(\mathbf{r}) = \alpha_0 + \frac{\alpha_1}{1 + {\exp(m(q_\sigma(\mathbf{r}) - n))}}
    \end{aligned}\,,
\end{equation}
where $\alpha_0$ and $\alpha_1$ are constants that will make the function switch between two different GGA functionals, and $m$ and $n$ are hyperparameters that determine the sharpness and shift in x-axis, respectively (see Fig.\ref{fig:alpha_deriv}). Specific values of these parameters are discussed in subsequent sections.

With this form defined, the full potential (Eq.~\ref{eq:convPBEalphamultipole}) can be evaluated using the derivative of $\alpha_\sigma$ with respect to $q_\sigma$:
\begin{equation}
    \frac{\dpa\alpha_\sigma}{\dpa q_\sigma} = \frac{-\alpha_1 m}{2 + \exp{-m (q_\sigma - n)} + \exp{m(q_\sigma - n)}}\,.
\end{equation}
As $m \rightarrow \infty$, the derivative of Eq.~\ref{eq:alphafunctionalform} becomes a $\delta$ function and can hence be neglected, since the width of function is 0. %The implications of this form of $\alpha$ and large values of $m$ are discussed in Sec. \ref{sec:XCpotential}.

\subsection{Trade-offs in accuracy for gases and solids with PBE$\alpha$}
The aim of this work is to show that the use of convolutional inputs to XC functionals facilitate the construction of functionals that go beyond the constraints of the GGA model space without significant additional expense. To achieve this, we seek to construct a functional that enables switching between different GGA approximations within a single simulation cell. This is achieved by allowing the $\alpha$ parameter of PBE$\alpha$ to vary in space, as originally proposed by Madsen,\cite{PhysRevB.75.195108} where we utilize a convolutionally derived monopole feature to determine the appropriate $\alpha$ value at a given point in space. To assess the performance of our functional for diverse electronic environments, we evaluated various properties such as the formation energy of molecules, the cohesive energy and lattice constants of metals, and the adsorption energy of small molecule adsorption systems, as described in Sec. \ref{subsec:data_gen}. 

\begin{figure}
\begin{center}
\includegraphics[width=8.5cm]{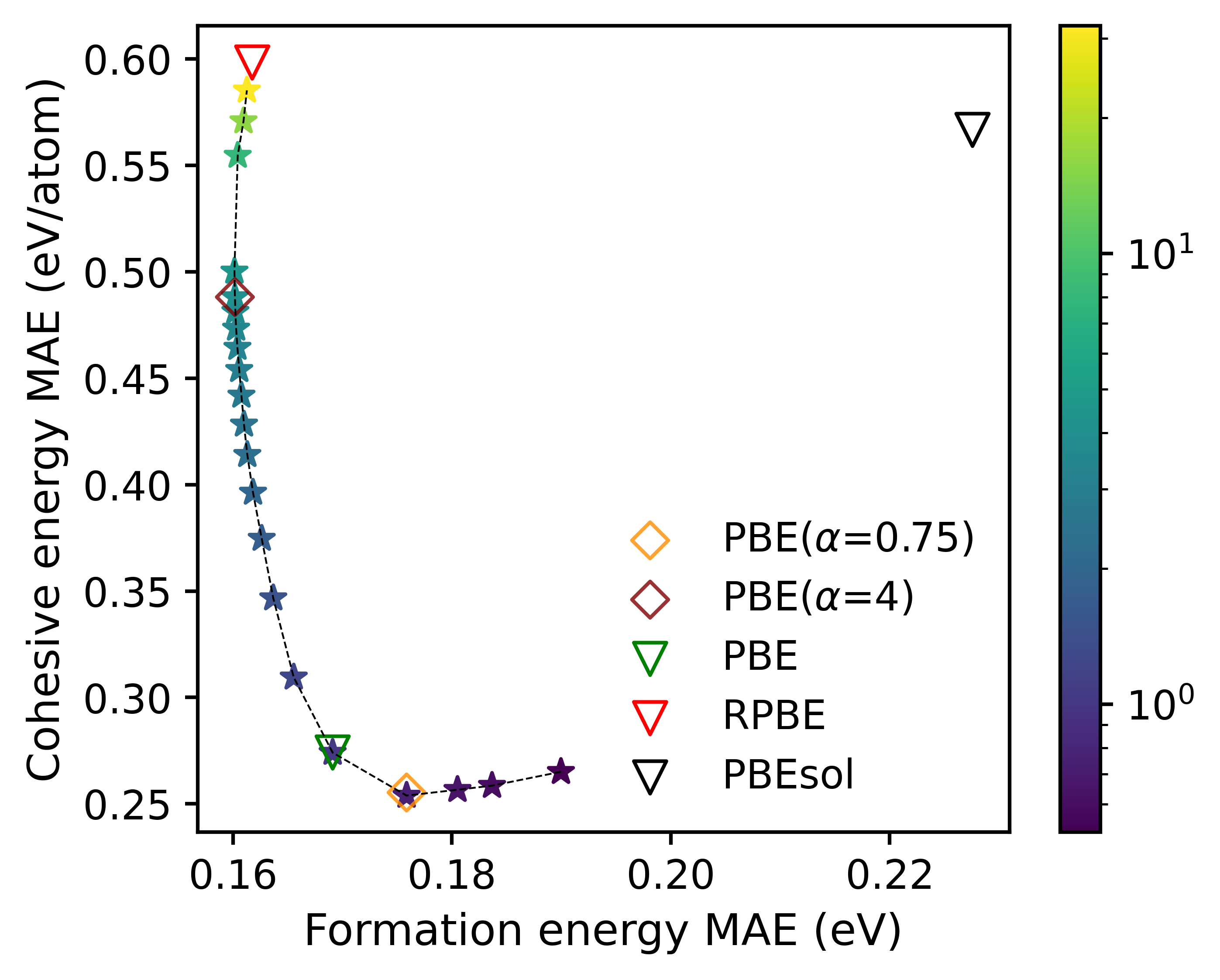}
\vspace*{-5mm}
\caption{Pareto plot for comparison of MAE in formation energy w.r.t CCSD(T) for molecules and MAE in the cohesive energy w.r.t experiment for metals. $\alpha$ is varied to different constant values from 0.52 to 32 to obtain a range of solutions for the metal and molecular datasets. The colorbar indicates range of values of $\alpha$. The MAE for PBE, RPBE, and PBEsol results are reported as well.}
\label{fig:const_alpha}
\end{center}
\end{figure}

Rather than directly fitting parameters in Eq.~\ref{eq:alphafunctionalform}, we utilize manual analysis to identify parameters intuitively. To determine $\alpha_0$ and $\alpha_1$, we find the optimum $\alpha$ that gives the lowest MAE for the formation energy of molecules compared to CCSD(T)\cite{cccbdb} and the lowest MAE for the cohesive energy of metals compared to experimental values \cite{BEEF-vdW}.
%the PBE($\alpha=0.52$) (which matches the slowly varying electron density limit defined by the WC functional).\cite{WCFunctional}
We select a range of values for $\alpha$ from 0.52 to 32. The MAE in cohesive energy is lowest when $\alpha = 0.75$, hence $\alpha_0$ is set as 0.75. The MAE in the formation energy of molecules is lowest when $\alpha = 4$, therefore, $\alpha_1$ is selected such that the maximum value of $\alpha$ does not exceed 4. The Pareto frontier for constant $\alpha$ functionals is shown in Fig. \ref{fig:const_alpha}. 
%Based on this observation, the functional form for $\alpha$, Eq.\ref{eq:alphafunctionalform}, was selected to interpolate between 0.75 and 4. 
Based on the fact that the electron density is generally larger for metallic environments than molecular environments, we select the 0$^{\mathrm{th}}$ order MP descriptor with a radial cut-off of 1.587 \AA ~(3 Bohr) as the input to the functional. We explored other options and found that this feature provided the best separation of metallic and molecular environments (see SI). \textcolor{black}{For simplicity, we use the $m \rightarrow \infty$ limit, which allows the additional convolutional term to be neglected and causes the functional to act as a discontinuous ``switch'' between $\alpha = 0.75$ and $\alpha = 4$ limits, such that at any point in space the functional should be either optimal for gas-phase or optimal for metals. We note that this discontinuity may introduce ambiguities in the ground state determined by the variational approach, but since the functional smoothly approaches the discontinuous limit (see SI), we expect these effects to be minor.} Finally, the parameter $n$ is selected to minimize errors on the molecular and metallic energies, as discussed in Sec. \ref{sec:tradeoff}.

% The selection of $m$ is based on evaluating the effect of the XC potential on the self-consistent energies, as discussed in Sec. \ref{sec:XCpotential}, and $n$ is selected to minimize errors on the molecular and metallic energies, as discussed in Sec. \ref{sec:tradeoff}.

\subsection{Trade-off between molecular formation and metal cohesion energies}
\label{sec:tradeoff}

\begin{figure*}[hbt!]
    \centering
    % left bottom right top
    \includegraphics[trim=95 110 90 120, keepaspectratio=true,scale=0.7,clip=True]{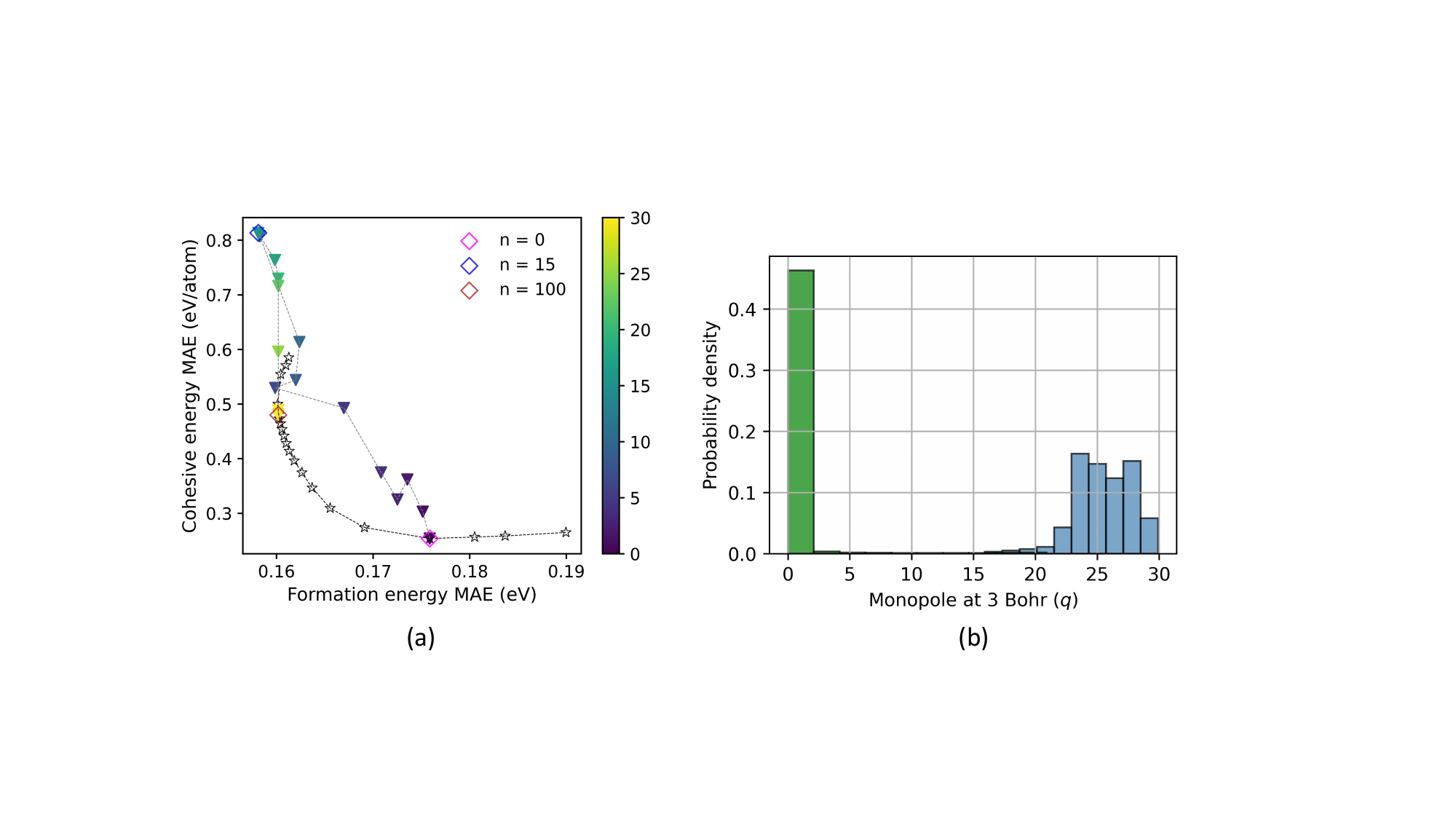}%wid\thetah=1.\linewid\thetah
    \caption{(a)Pareto plot for comparison of MAE in formation energy w.r.t CCSD(T) for molecules and MAE in cohesive energy w.r.t WC for metals for multipole functionals where $n$ is varied from 0 to 100 to obtain a range of solutions. \textcolor{black}{The stars in black show when $\alpha$ is varied to different constant values from 0.52 to 32.} The colorbar indicates range of values for parameter n in the multipole functional. (b) Histograms for Rh atom in a box (green) and bulk Rh cubic cell (steelgray). Cohesive energy necessitates spin-polarized calculations for single metal atoms. Therefore, in the histogram, we visualize the average value of monopole feature for spin-up and spin-down densities.}\label{fig:tradeoff}
\end{figure*}

To determine the value of $n$, the final parameter of Eq.~\ref{eq:alphafunctionalform}, we evaluate the performance of the functional for two metrics: the molecular formation energy and metal cohesion energy.
In Fig. \ref{fig:tradeoff} (a), we present the performance of the PBEq functional for $m=10^6$  with $n$ ranging from 0 to 100, where $n=0$ is equivalent to PBE($\alpha=0.75$) and $n=100$ is equivalent to PBE($\alpha=4$) since $0 < q(\mathbf{r}) < 100   \, \forall  \, \mathbf{r}$. The results show that the multipole functional with $n = 15$ slightly improves the molecular formation energy compared to RPBE with MAE = 0.158 eV. However, for cohesive energy, the functional does not demonstrate improved performance for any value of $n$. This lack of improvement is rationalized by the fact that the cohesive energy is computed from both a solid-state quantity (energy of the bulk metal) and a gas-phase quantity (energy of the metal atom in vacuum), and thus cancellation of error may not occur in cases where the gas-phase metal contains environments that cause $\alpha$ to tend toward the $\alpha=4$ limit, but the bulk metal contains environments where $\alpha$ tends toward $\alpha=0.75$. Indeed, the histogram of the monopole features for Rh (Fig. \ref{fig:tradeoff} (b)) reveals that the single atom has monopole magnitudes closer to zero, and will hence be treated like a gas-phase molecule. In this case, cancellation of error will not occur, and we hypothesize that the relatively poor cohesive energies are due to incorrect treatment of the metal atom in the gas phase. Despite the relatively poor performance on cohesive energies, the $n=15$ functional is outside the Pareto frontier of PBE$\alpha$ due to its improved performance on molecular formation energies, and we set $n=15$. Subsequently, we will refer to this functional as PBEq15.

\begin{figure}
\begin{center}
\includegraphics[width=8.4cm]{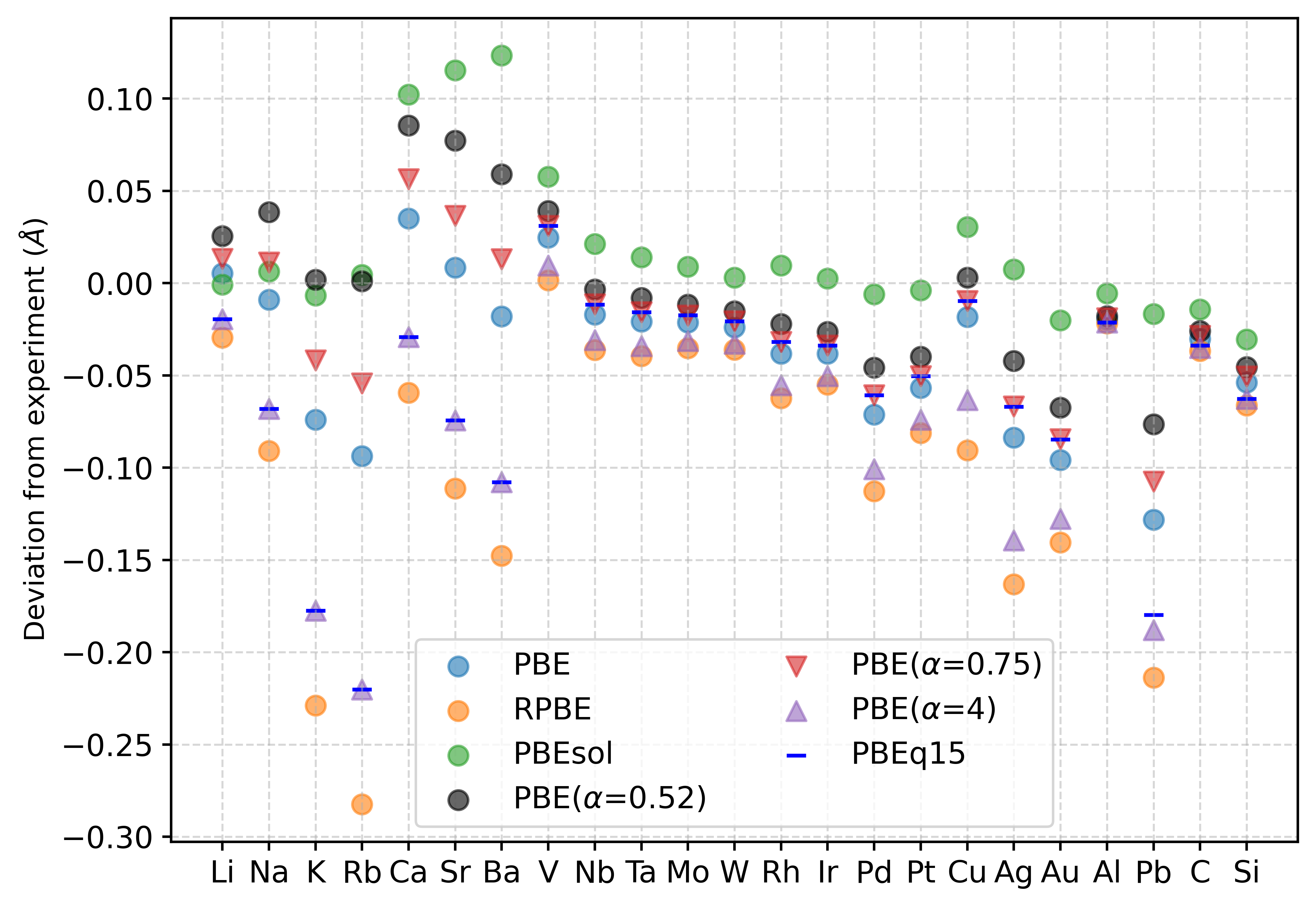}
\vspace*{-10mm}
\caption{Deviation of lattice constant from experimental value for different functionals calculated using the equation of state approach.}
\label{fig:lattice_constants}
\end{center}
\end{figure}

In the case of metal cohesive energies, similar behavior is observed even with functionals specifically designed for solids such as PBEsol (Fig. \ref{fig:const_alpha}). To further corroborate the validity of the treatment of solid-state systems, we use the PBEq15 functional to calculate the lattice constant of bulk metals, a quantity that only involves solid-state electronic environments. The results, shown in Tab. \ref{tab:lc_comp}, indicate that the lattice constants of PBEq15 exhibit mean errors that are in between PBE($\alpha$=0.75) and PBE($\alpha$=4). A more detailed comparison, shown in Fig. \ref{fig:lattice_constants}, reveals that each individual lattice constant is bounded by the $\alpha=0.75$ and $\alpha=4$ limit, with $d$-block metals and others with more valence electrons approaching the $\alpha=0.75$ limit as expected based on the higher electron density and corresponding higher monopole values. These findings corroborate that the treatment of solids is generally improved over RPBE and PBE($\alpha$=4). We further evaluate the performance of this functional for representative systems of gas adsorption on metals in the subsequent section, which provides a test of error cancellation in a practical scenario where both molecular and metallic environments are present.

\begin{table}
	\begin{center}
	\caption{Comparison of MAE for lattice constants obtained from different XC functionals.}\label{tab:lc_comp}
		\begin{tabular}{lcc}	
\toprule		
Functional & MAE [\AA]\\
\midrule
PBE 	&	0.043\\
RPBE 	&   0.093\\
PBEsol  &    0.026\\
PBE($\alpha=0.52$)	& 0.034\\
PBE($\alpha=0.75$)	& 0.037\\
PBE($\alpha=4$) & 0.076\\
PBEq15 & 0.062\\
\bottomrule	
    \end{tabular}
        \end{center}
%\footnotesize{\textsf{[a] Table footnote. [b] Table footnote. [c] Table footnote.}}

\end{table}

\subsection{Performance of PBEq15 functional for adsorption energies}

Adsorption energies are used to predict the performance of solid catalysts as they provide insight into the strength of surface-adsorbate bonds in thermal and electrochemical reactions.~\cite{abild2007scaling,fernandez2008scaling,hammer1996co,medford2015sabatier,fields2017scaling,christoffersen2001anode,lopez2001synergetic,RPBE,WELLENDORFF201536} DFT is widely employed for these adsorption energy calculations to gain a deeper understanding of surface-adsorbate phenomena.~\cite{norskov2011density,nilsson2011chemical,nilsson2005electronic,nilsson2004chemical} The existing scientific literature suggests that local or semilocal functionals such as LDA or PBE tend to overestimate chemisorption energies for surface systems.~\cite{jones1989density,RPBE,grinberg2002co,gil2003site} A more specific qualitative failure of most known \textcolor{black}{GGA} functionals is the well-known ``CO puzzle'' where \textcolor{black}{calculations} suggests that CO exhibits a preference for \textcolor{black}{high}-coordination sites, contrary to findings from many low-temperature experimental studies such as scanning-probe, diffraction and spectroscopic observations \textcolor{black}{that show preference for low-coordination sites.}~\cite{feibelman2001co} This problem can be attributed to overestimation of the bond strength between the empty molecular frontier orbital and the substrate.

\begin{figure*}
    \centering
    \includegraphics[keepaspectratio=true,scale=0.55,clip=True]{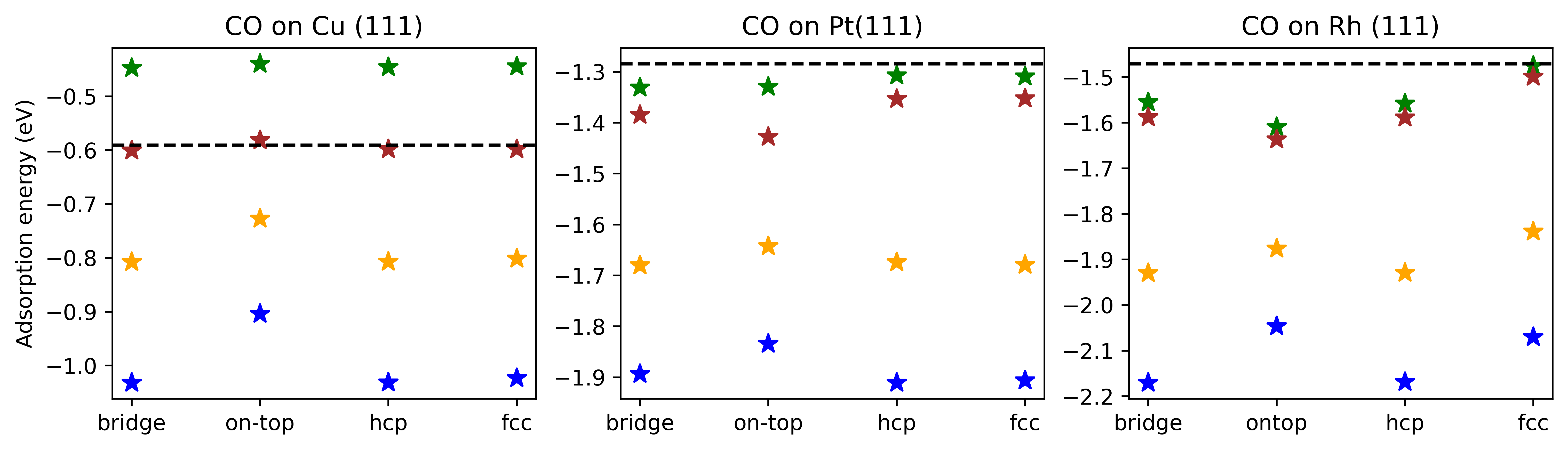}%wid\thetah=1.\linewid\thetah
    \caption{CO adsorption energy on (a) Cu(111), (b) Pt(111) and (c) Rh(111) on four sites: bridge, on-top, hcp and fcc. Results from experiment\cite{WELLENDORFF201536} and DMC are included where available.}\label{fig:adsorption_energy}
\end{figure*}

To investigate the performance of our functional on adsorption energies, we focus on three common cases involving CO adsorption: Cu (111), Pt(111), and Rh(111) for which GGA functionals predict the wrong site order for adsorption.~\cite{schimka2010accurate} For these calculations, we compare the PBEq15 functional with RPBE, PBE, and PBE($\alpha=0.52$) for all high-symmetry adsorption sites on each metal. The evaluation of different binding sites provides insight into cancellation of error under different environments, and enables a qualitative assessment of whether a given functional can predict the experimentally-known adsorption site. However, the comparison to experiment is complicated by the fact that the exact experimental geometry is unknown. For this reason, we further supplement the comparison with DMC results in the case of Pt(111), providing a comparison at identical atomic positions. Together, these evaluations help provide a detailed picture of how the PBEq15 functional performs in these representative scenarios. 

Fig. \ref{fig:adsorption_energy} illustrates the CO adsorption energy on four different sites on the (111) surfaces and shows that the PBEq15 functional yields adsorption energies between RPBE and PBE, suggesting that error cancellation is occurring. Analysis of Fig. \ref{fig:adsorption_energy} reveals several observations. First, the calculated adsorption energy from the PBEq15 functional is close to the experimental value for all three metal surfaces. In particular, the functional demonstrates excellent agreement with the experimental adsorption energy for Cu. Second, the PBEq15 results are the closest to the DMC results for the on-top site on Pt(111). The DMC results are systematically higher than all functionals for other adsorption sites, with RPBE results being slightly closer to DMC (but also close to PBEq15). Finally, the PBEq15 functional qualitatively predicts the correct binding site (on-top) for CO on Pt, a task that is generally challenging for semi-local functionals.~\cite{feibelman2001co, grinberg2002co} The DMC results confirm that the on-top site should be most stable for the atomic geometries used here, with the relative error bars on DMC results showing a clear preference for the on-top site (see Fig.~\ref{supp:fig:adsorption_energy_err} in SI).
%To our knowledge, this is the only semilocal functional that correctly predicts the CO binding site on Pt(111). 
Given the relatively small number of systems studied, we recognize that these successes may be fortuitous, and more expansive testing is outside the scope of this study. However, these observations suggest that despite the poor performance of the PBEq15 functional for cohesive energy, cancellation of error is valid for adsorption energy calculations. These findings indicate that the strategy of using advanced electron density features to ``switch'' between functionals in a single calculation may prove fruitful for surface science functionals.

\begin{figure}
\begin{center}
\includegraphics[width=8cm]{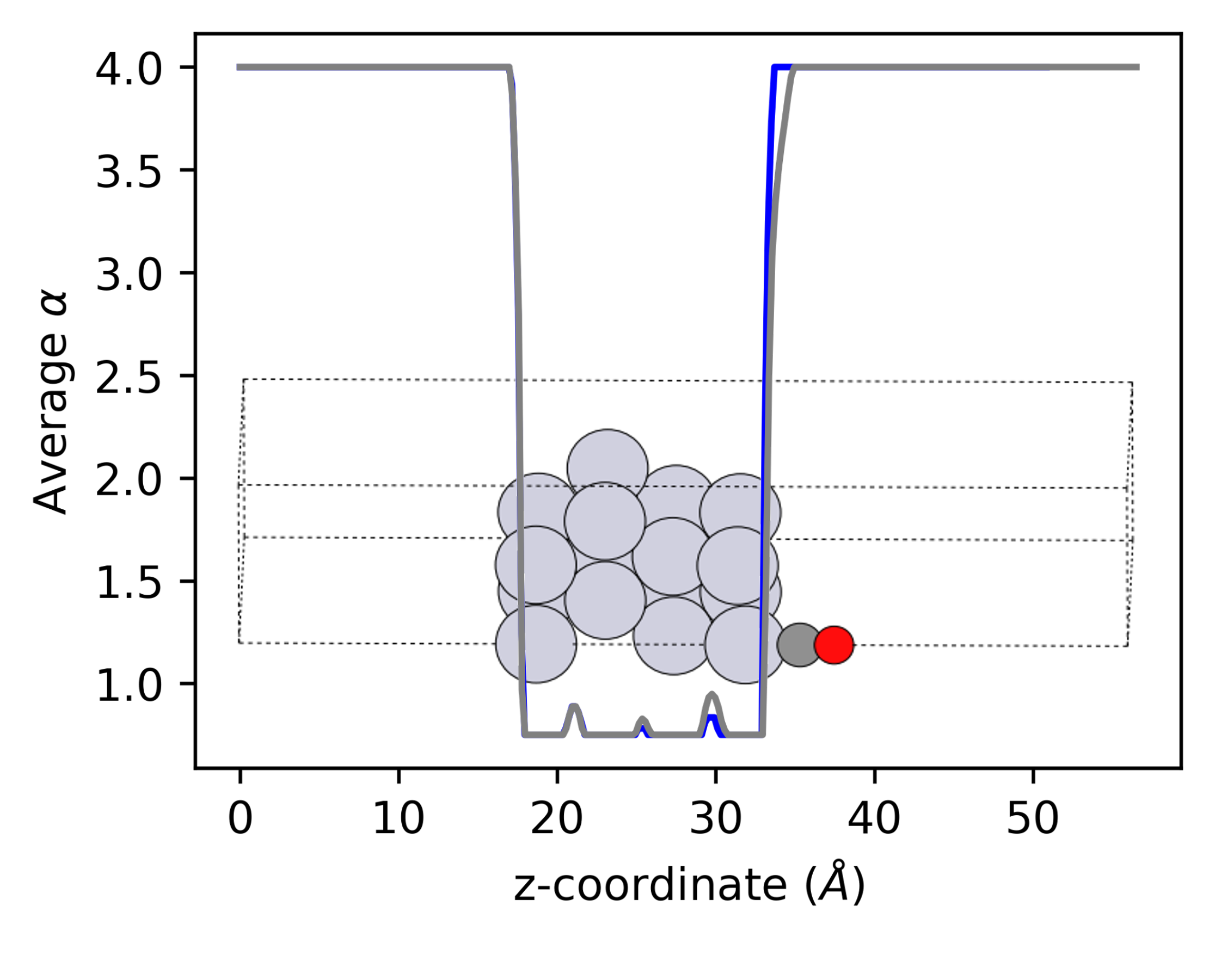}
\caption{Mean $\alpha$ versus z-coordinate for slab (gray) and slab + adsorbate (blue) system providing a visual representation of PBEq15 functional transition from lower to higher values of $\alpha$ in metallic and molecular environments, respectively.}
\label{fig:slab_avg_alpha}
\end{center}
\end{figure}

To further analyze the behavior of the PBEq15 functional, we evaluate the spatial distribution of $\alpha$ for a surface system.
Fig. \ref{fig:slab_avg_alpha} illustrates the average $\alpha$ as a function of the z-coordinate for Pt(111) and CO adsorbed on Pt(111). The $\alpha$ values for both the slab and the chemisorbed system are similar. Furthermore, it is clear that the functional transitions between lower and higher values of $\alpha$ in the metallic and molecular electronic environments, respectively. These observations confirm that the functional behaves intuitively and corroborate the cancellation of error between the slab system with and without the adsorbate present.

% \begin{table*}
% 	\begin{center}
% 	\caption{Caption for a two-column table.}\label{tab2}
% 		\begin{tabular}{lccccc}	
% \toprule		
% Head 1\textsuperscript{[a]} & Head 2 [Unit] & Head 3\textsuperscript{[b]} [Unit] & Head 4 [Unit] & Head 5 & Head 6\\
% \midrule
% Entry 1 	&	 Value 1	&	Value 2	&	Value 3	&	Value 4	&	Value 5\\
% Entry 2 	&	 Value 1	&	Value 2	&	Value 3	&	Value 4	&	Value 5\\
% Entry 3\textsuperscript{[c]} 	&	 Value 1	&	Value 2	&	Value 3	&	Value 4	&	Value 5\\
% Entry 4 	&	 Value 1	&	Value 2	&	Value 3	&	Value 4	&	Value 5\\
% \bottomrule	
% 	\end{tabular}
% 	\end{center}
% \footnotesize{\textsf{[a] Table footnote. [b] Table footnote. [c] Table footnote.}}

% \end{table*}

\section{Conclusion}
\label{conclusion}
	
In this study, we derive the variational derivative of functionals constructed using inputs obtained from convolutions, providing a route to obtaining the exchange-correlation potential of these functionals.
%present a general derivation obtaining the potential for functionals that use arbitrary convolutional kernels to generate inputs. 
In addition, we present and implement a proof-of-concept functional, PBEq, that can transition between molecular and metallic electronic environments. This functional is dependent on a physical parameter, denoted as $\alpha$ which varies in space depending on the local monopole of the electron density. We assess its performance by evaluating the formation energy of molecules, the cohesive energy and lattice constants of metals, and adsorption energy of small surface systems, including a comparison to high-accuracy DMC calculations of CO adsorbed on Pt(111). The results indicate that even a relatively simple functional with a few parameters that are set based on intuition and manual analysis is capable of performing as well as or better than existing GGA functionals for molecular formation energies and the adsorption of small molecules on metals.

The results also indicate several key lessons for future efforts. We show that even for energy functionals that are bounded by known physical limits, additional terms in the XC potential can induce density errors that lead to unphysical behavior. This indicates that future machine-learned energy functionals should be carefully constructed or fitted directly to XC potential information to avoid spurious density errors in self-consistent calculations. Furthermore, the importance of error cancellation is highlighted by the fact that cohesive energies do not improve even when the treatment of the bulk metal system does. This emphasizes the importance of carefully selecting target properties for optimization and ensuring that functionals are sufficiently smooth to allow error cancellation in practical scenarios. 

The general framework outlined in this work lays the foundation for a wide range of new functional design strategies. Convolutional neural networks could be used to directly learn the connection between energy and electron density, and the equations herein provide a straightforward way of computing the corresponding potential. Alternatively, convolutions can be used to calculate descriptors, such as MP features or others, and these descriptors can be used to replace or expand the model spaces defined by Jacob's ladder. Physically derived functional forms, machine learning models, or hybrid models can be used to identify and construct functionals that are based on these descriptors, providing both data-driven and physically inspired routes for constructing new functionals. We note that these possibilities are not limited to XC functionals, since the derivations are general to any functional, and may also find application in the design of density-dependent kinetic energy functionals, such as those used in orbital-free DFT or functionals in other areas of science which are associated with a variational problem. We hope that the results presented here inspire the community to creatively explore the possibilities and create a range of new density functionals with improved performance across the spectrum of quantities of interest.

\section{Methods}
\label{sec:methods}

\subsection{Computational details}
We implement the self-consistent multipole-dependent PBEq functional in the SPARC software--- a real-space DFT code that has comparable accuracy to established planewave codes, while requiring walltimes that are more than an order of magnitude lower.\cite{ghosh2017sparc1, Xu2021Sparc:Calculations,  zhang2023version} In all calculations, we choose optimized norm-conserving Vanderbilt (ONCV) pseudopotentials~\cite{Hamann2013OptimizedPseudopotentials} from the SPMS~\cite{SHOJAEI2023108594} collection. The PBEq functional implementation is employed for periodic cells, hence for all DFT calculations, we use periodic boundary conditions. All parameters have been chosen to provide numerical accuracy of at least 0.027 eV/atom (0.001 Ha/atom) in the energy. The atomic positions and structures are defined using the Atomic Simulations Environment (ASE) package.~\cite{Khorshidi2016Amp:Simulations} Note that we only modified the exchange energy functional and used the standard PBE correlation functional in our implementation. All calculations are spin-paired except for the single metal in a box calculation to obtain the cohesive energies. For the spin-polarized calculations, we fix the initial guess for orbitals and additional eigensolve iterations to improve the stability of single metal atoms. We utilize fixed atomic positions for all calculations since forces and stresses are not yet implemented for the PBEq functional. \textcolor{black}{The current SPARC implementation of MP features supports two radial probes, Heaviside and Legendre polynomials. In this work, we utilize a Heaviside function as the radial probe, which is conceptually simplest since it corresponds to a hard cutoff. The convolutions are evaluated using real-space integration within the code, and energies calculated with these descriptors are converged to a numerical error below 1e-3 Ha/atom (see SI).} The code and examples of input files along with instructions necessary for reproducing the results of this work can be found on Zenodo (\url{https://doi.org/10.5281/zenodo.10660363}).

We perform all QMC calculations using the QMCPACK software\cite{qmcpack} and the Nexus workflow management system.\cite{nexus} Using single-particle orbitals from DFT calculations with the RPBE functional, we construct Slater-Jastrow type trial wave functions of the form $\Psi(\mathbf{R}) = D^{\uparrow}D^{\downarrow}e^J$. Here, $D^\uparrow$ and $D^\downarrow$ are the Slater determinants consisting of up- and down-spin single-particle orbitals, respectively, and $e^J$ is the Jastrow factor, which incorporates electron correlation into the wave function ansatz beyond the mean-field level. We include one-, two-, and three-body Jastrow factors in our calculations. Following an initial optimization of the Jastrow factors and the trial wave function using variational Monte Carlo simulations,\cite{foulkes_qmc} we perform DMC calculations with a timestep of 0.01 a.u. In each case, our walkers are set to fluctuate between 12,000 and 24,000 DMC walkers to avoid walker population bias. In DMC and other many-body electronic structure calculations of periodic solids, it becomes necessary to treat many-body finite-size effects, in addition to the typical one-body finite-size effects that arise in DFT, due to the use of periodic boundary conditions.\cite{drummond_finite_size} First, to eliminate one-body finite-size effects, we employ twist-averaging (analogous to k-point grid sampling in DFT) in our QMC calculations.\cite{twist_averaging} For the primitive cell, we use an $8\times8\times1$ twist grid. Many-body finite-size errors can only be mitigated through tiled replication of the primitive cell along the periodic dimensions (parallel to the slab in this case). However, it has been shown previously that the calculation of binding energies in chemisorption systems such as that modeled here is accompanied by the cancellation of many-body finite-size errors, which results in well-converged binding energies even at small cell sizes.\cite{iyer_rubenstein_dmc} Therefore, we use only the primitive cell, with an $8\times8\times1$ twist grid, for DMC binding energy calculations.

\subsection{Data generation}\label{subsec:data_gen}

We have utilized various datasets containing both structural and energetic information, encompassing a diverse range of electronic structures ranging from molecular to metallic environments. These datasets have been sourced from existing literature and are summarized briefly as follows. Additional information about the datasets can be found in the SI.
\begin{itemize}
    \item[(a)] \textit{Molecular formation energies}: A small set consisting of geometries and energies of 217 small organic molecules made of H, C, N and O atoms are taken from the Computational Chemistry Comparison and Benchmark Database (CCCBDB). These calculations are done at CCSD(T)-ccPVTZ level of theory.~\cite{cccbdb} A linear correction scheme is applied to the raw DFT and CCSD(T) energies to compare the formation energies across different levels of theory. The formation energy, $E_f$ is given by:
    \begin{equation}
        \begin{aligned}
            E_{f} = E_{\mathrm{total}} - \sum_{i=1}^4 n_i E_i
        \end{aligned}
    \end{equation}

    where $n_i$ is the number of atoms for each element type (H, C, N and O) in the molecule. $E_i$ are the coefficients for each element type fitted using the referencing scheme. 
    \item[(b)] \textit{Solid state properties}: The solid state reference data used in this study is obtained from data curated by Wellendorff et al., which contains lattice constants and cohesive energies of cubic lattices.~\cite{BEEF-vdW} We only consider spin-paired systems for these calculations. For these calculations, we only focus on spin-paired systems for bulk lattices. 
    The cohesive energy, $E_C$ for a transition metal bulk system is given by:
    \begin{equation}
        \begin{aligned}
            E_{C} = E_{\mathrm{single\,atom}} - E_{\mathrm{bulk}}/n_{\mathrm{atoms}}
        \end{aligned}
    \end{equation}
    where $E_{\mathrm{single\,atom}}$ is the spin-polarized energy of a single metal atom in a box, $E_{\mathrm{bulk}}$ is total energy of cubic cell for the bulk and $n_\mathrm{atoms}$ is the total number of atoms in the primitive cell.

    %Lattice constants were optimized by calculating energies within +/- ??? \% of the experimental value and fitting a cubic equation of state. The energy of the minimum lattice constant is then computed and the equation of state is re-fitted. This procedure is repeated until the equation of state minimum value is within 0.005 Å of a DFT-calculated value.
    Lattice constants were determined using an equation of state approach. First, 10 lattice constants within +/- 10\% of the PBE values from Wellendorf et. al\cite{BEEF-vdW} were computed, and the optimum was determined by fitting a cubic polynomial equation of state. Then, the results were refined using 20 lattice constants within +/- 4 \% of the optimum. The equation of state is finally re-fitted to these results to determine the optimum, and in all cases the equation of state minimum value is within 0.01 \AA ~of a DFT calculated value.

    \item[(c)] \textit{Adsorption energies}: For chemisorption, we select CO adsorbed on three transition metal surfaces: Pt(111), Rh(111) and Cu(111). The surfaces are constructed using PBE lattice constants obtained from a benchmark dataset for transition metal surfaces by Wellendorf et. al at low coverage.~\cite{WELLENDORFF201536} The adsorption energy, $E_{\mathrm{adsorption}}$ for a system is given by:
    \begin{equation}
        \begin{aligned}
            E_{\mathrm{adsorption}} = E_{\mathrm{slab,molecule}} - E_{\mathrm{slab}} - E_{\mathrm{molecule}}
        \end{aligned}
    \end{equation}
\end{itemize}

%\clearpage

\section*{Acknowledgements}

S.J.S was a recipient of the Novelis Graduate Scholarship during the period this research conducted. The authors acknowledge the funding provided by the U.S. Department of Energy, Basic Energy Sciences, Computational Chemical Sciences Program, through grant numbers DESC0019410 and DESC0019441. This research was also supported by the supercomputing infrastructure provided by Partnership for an Advanced Computing Environment (PACE) through its Hive (U.S. National Science Foundation (NSF) through grant MRI-1828187) and Phoenix clusters at Georgia Institute of Technology, Atlanta, Georgia. The views and conclusions contained in this document are those of the authors and should not be interpreted as representing the official policies, either expressed or implied, of the Department of Energy, or the U.S. Government. We thank Kyle Bystrom for pointing out the issues related to density scaling and providing useful references. S.J.S. thanks Gabriel S. Gusmão for helpful discussions about the functional forms of $\alpha$ and their derivatives.

%The authors acknowledge the funding provided by the U.S. Department of Energy, Basic Energy Sciences, Computational Chemical Sciences Program under grant number DESC0019410.
\section*{Conflict of Interest}

The authors declare no competing financial interest.

%%%%%%%%%%%%%%%%%%%%%%%%%%%%%%%%%%%%%%%%%%%%%%%%%%%%%%%%%%
%%%%%%%%%%%%%%%%%%%%%%%%%%%%%%%%%%%%%%%%%%%%%%%%%%%%%%%%%%
%%%%%%%%%%%%%%%%%%%%%%%%%%%%%%%%%%%%%%%%%%%%%%%%%%%%%%%%%%
\begin{shaded}
\noindent\textsf{\textbf{Keywords:} \keywords} 
\end{shaded}
%%%%%%%%%%%%%%%%%%%%%%%%%%%%%%%%%%%%%%%%%%%%%%%%%%%%%%%%%%
%%%%%%%%%%%%%%%%%%%%%%%%%%%%%%%%%%%%%%%%%%%%%%%%%%%%%%%%%%
%%%%%%%%%%%%%%%%%%%%%%%%%%%%%%%%%%%%%%%%%%%%%%%%%%%%%%%%%%

%%%%%%%		References			%%%%%%% 

\setlength{\bibsep}{0.0cm}
\bibliographystyle{Wiley-chemistry}
\bibliography{example_refs}

\clearpage

%%%%%%%		TOC Entry			%%%%%%% 

\section*{Entry for the Table of Contents}

%	 please select one option only and delete the other one

%%%%%%%		Option 1			%%%%%%%    

\noindent\rule{11cm}{2pt}
\begin{minipage}{5.5cm}
\includegraphics[width=5.5cm]{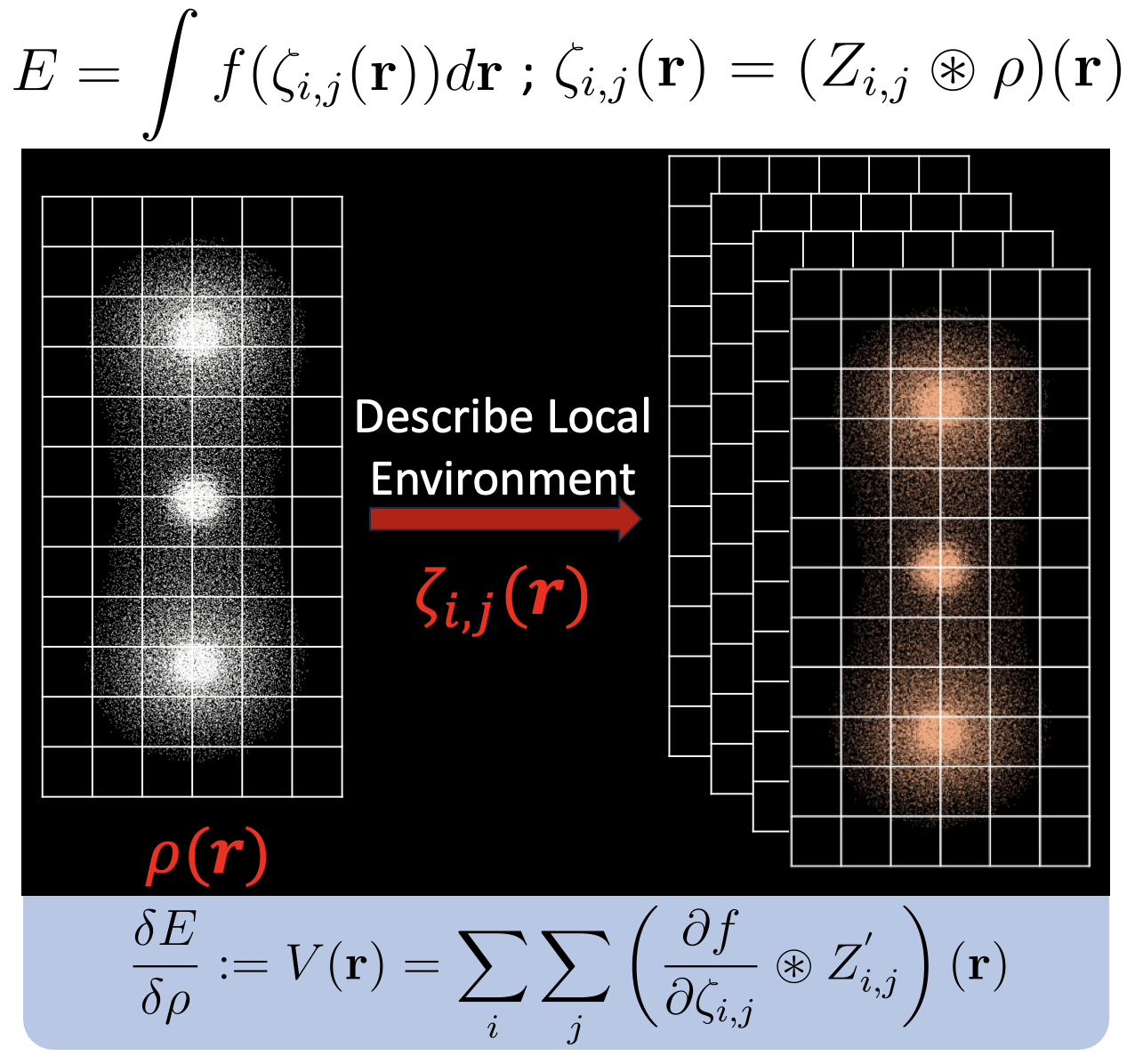}
\end{minipage}
%\hspace{0.5cm}
\begin{minipage}{5.5cm}
\large\textsf{PBEq class of self-consistent XC functionals constructed from convolution of electron density with arbitrary kernels. This functional can distinguish between electronic environments within a system by employing different approximations at different points in space.}
\end{minipage}
\noindent\rule{11cm}{2pt}

\vspace{2cm}

%%%%%%%		 Option 2			%%%%%%%    

% \noindent\rule{11cm}{2pt}
% \begin{minipage}{11cm}
% \includegraphics[width=11cm]{latex/graphical_abstract.png}
% \end{minipage}
% \begin{minipage}{11cm}
% \large\textsf{PBEq class of self-consistent XC functionals constructed from convolution of electron density with arbitrary kernels. This functional employs different approximations at different points in space within a given system and distinguishes between electronic environments.}
% \end{minipage}
% \noindent\rule{11cm}{2pt}

%%%%%%%%%%%%%%%%%%%%%%%%%%%%%%%%%%%%%%%%%%%%%%%%%%%%%%%%%
%%%%%%%%%%%%%%%%%%%%%%%%%%%%%%%%%%%%%%%%%%%%%%%%%%%%%%%%%
%%%%%%%%%%%%%%%%%%%%%%%%%%%%%%%%%%%%%%%%%%%%%%%%%%%%%%%%%

\end{document}

\maketitle
%\myexternaldocument{chemistry-template}

\section{GGA exchange potential as variational derivative of convolutional density functional approximation}\label{sec:pbe_derivation}

In case of GGA, the exchange  energy is written as a function of the $\edens$ and its gradient, $\grad\edens$:
\begin{equation}
     E[\rho] = \anaint{f(\edens, \grad \edens)}
\end{equation}

The ``descriptors'' for this particular case are $\edens$ and $\grad \edens$. For clarity, the gradient vector, $\grad \edens$ can be expanded into three scalar quantities, $\frac{\dpa \edens}{\dpa 1}$, $\frac{\dpa \edens}{\dpa 2}$, and $\frac{\dpa \edens}{\dpa 3}$ that represent the x, y, and z directions, respectively.  Under certain scenarios, such as when the electron density is represented on a finite difference grid with periodic boundary conditions, the inputs to $f$ can be interpreted as convolutions:
%In this case, all descriptors can be directly interpreted as convolutions:

\begin{equation}
\begin{aligned}
        \zeta_{0,0} &= \edens = \conv{\delta}{\rho} \,,\\
         \zeta_{1,0} &= \conv{\nabla_1}{\rho} \approx \frac{\dpa}{\dpa x} \edens  \,, \\
         \zeta_{1,1} &= \conv{\nabla_2}{\rho} \approx \frac{\dpa}{\dpa y} \edens  \,, \\
         \zeta_{1,2} &= \conv{\nabla_3}{\rho} \approx \frac{\dpa}{\dpa z}  \edens  \,, \\
\end{aligned}
\end{equation}
where $\zeta_{i,j}$ represents output of convolution of electron density, $\edens$ with kernels $\delta$, $\nabla_1$, $\nabla_2$ and $\nabla_3$. $\delta$ is the delta-function and $\nabla$ operator denotes a numerical 
%12$^{th}$ order 
finite-difference stencil for the gradient along the $x$, $y$ or $z$ directions, and the $\approx$ symbol denotes the fact that there will be numerical error due to a finite grid spacing and finite order of the derivative stencil. The convolutional kernels \stencils{} in this scenario are $Z_{0,0} = \delta$,  $Z_{1,0} = \nabla_1$,  $Z_{1,1} = \nabla_2$, and $Z_{1,2} = \nabla_3$, and $Z'_{0,0} = \delta$,  $Z'_{1,0} = \nabla_{-1} = -\nabla_1$,  $Z'_{1,1} = \nabla_{-2} = -\nabla_2$, and $Z'_{1,2} = \nabla_{-3} = -\nabla_3$, where the sign change arises in the $\nabla$ stencils since they are anti-symmetric. Following the general derivation from main text and substituting in Eq. \ref{main:eq:v_potential}:
\begin{equation}
\begin{aligned}
        V(\mathbf{r}) &= \conv{\frac{\dpa f}{\dpa \edens}}{\delta} +  \sum_{j=x,y,z} \conv{\frac{\dpa f}{\dpa \nabla_j \edens}}{(-\nabla_j)} \\
        &= \frac{\dpa f}{\dpa \edens} - \sum_{j=1}^{3} \conv{\nabla_j}{\frac{\dpa f}{\dpa \nabla_j \edens}}
\end{aligned}
\end{equation}

which is consistent with the well-known result:

\begin{equation}
    V(\vec{r}) = \frac{\dpa f}{\dpa \edens} - \grad \cdot \frac{\dpa f}{\dpa \grad \edens}
\end{equation}

where the divergence can be expanded as:

\begin{equation}
    \grad \cdot \frac{\dpa f}{\dpa \grad \edens} = \frac{\dpa}{\dpa x}\left( \frac{\dpa f}{\dpa \nabla_1 \edens} \right) + \frac{\dpa}{\dpa y}\left( \frac{\dpa f}{\dpa \nabla_2 \edens} \right) + \frac{\dpa}{\dpa z}\left( \frac{\dpa f}{\dpa \nabla_3 \edens} \right)
\end{equation}

\section{Full derivation of PBE$\alpha$ exchange potential as variational derivative of convolutional density functional approximation}\label{sec:pbealpha_derivation}

As described in the main text, the PBE$\alpha$ is a generalized form of the widely-used PBE functional. 
%The formulation of PBE$\alpha$ differs slightly from the general formulation of PBE presented above. 
The exchange energy density per particle is  a function of
%the product of $(\mathbf{r}))$, where 
$s_\sigma(\mathbf{r})$, which is the ``reduced'' gradient given by:

\begin{equation}
    \label{eq:s_sgima}
    s_{\sigma}(\mathbf{r}) = \frac{| \grad \rho_{\sigma}(\mathbf{r})|}{2 (6\pi^2)^{1/3} \rho_{\sigma}(\mathbf{r})^{4/3}}\,,
\end{equation}

where $| \grad \rho_{\sigma}(\mathbf{r})|$ is the $L_2$-norm of the gradient vector: $| \grad \rho_{\sigma}(\mathbf{r})| = \sqrt{\sum_j (\nabla_j \rho_{\sigma}(\mathbf{r}))^2}$.  For simplicity, we represent $\varepsilon_{x}^{GGA}(\spinedens,s_\sigma(\mathbf{r}))$ as:
\begin{equation}
    \varepsilon_{x,\sigma}^{GGA} = \rho_\sigma\varepsilon_{x,\sigma} F_{x,\sigma}\,,
\end{equation}

where $\varepsilon_{x,\sigma}$ is the LDA exchange energy density per particle given by:
\begin{equation}
    \begin{aligned}
        \varepsilon_{x,\sigma} = \frac{-3}{4\pi}(6\pi^2\rho_\sigma)^{1/3}\,,
    \end{aligned}
\end{equation}

$F_{x,\sigma}$ in PBE$\alpha$ is given by:

\begin{equation}
    \begin{aligned}
        F_{x\sigma}^{PBE_\alpha} = 1 + \kappa\left[1 - \frac{1}{\left(1+ \frac{\mu s_\sigma^2}{\kappa\alpha}\right)^\alpha}\right]\,.
    \end{aligned}
\end{equation}

The spin-dependent exchange potential for PBE$\alpha$, $V_{x,\sigma}^{PBE\alpha}$ is given by:

\begin{equation}
\begin{aligned}
    \label{eq:potentialPBEalpha}
     V_{x,\sigma}^{PBE\alpha} = \frac{\dpa\varepsilon_{x,\sigma}^{PBE\alpha}}{\dpa \rho_\sigma} - \sum_{j=1}^{3} \shortconv{\nabla_j}{\frac{\dpa  \varepsilon_{x,\sigma}^{PBE\alpha}}{\dpa\nabla_j \rho_\sigma}}\,,
\end{aligned}
\end{equation}

where $\frac{\dpa\varepsilon_{x,\sigma}^{PBE\alpha}}{\dpa \rho_\sigma}$ is given by:

\begin{equation}
    \begin{aligned}
    \label{eq:dparho}
        \frac{\dpa\varepsilon_{x,\sigma}^{PBE\alpha}}{\dpa \rho_\sigma} &= \frac{\dpa}{\dpa \rho_\sigma}\left(\rho_\sigma\varepsilon_{x,\sigma} F_{x,\sigma}^{PBE_\alpha}\right)\\
        &= \frac{\dpa}{\dpa \rho_\sigma}\left(\rho_\sigma \cdot \left( \frac{-3}{4\pi}(6\pi^2\rho_\sigma)^{1/3}\right)\cdot\left(1 + \kappa\left[1 - \frac{1}{\left(1+ \frac{\mu s_\sigma^2}{\kappa\alpha}\right)^\alpha}\right]\right)\right)\\
        &= \varepsilon_{x,\sigma}\left(\frac{4}{3}F_{x\sigma} +\mu\left(1+\frac{\mu s_\sigma^2}{\kappa\alpha}\right)^{-\alpha - 1}\left(\frac{-8s_\sigma^2}{3}\right) \right)\,,
    \end{aligned}
\end{equation}
and $\frac{\dpa  \varepsilon_{x,\sigma}^{PBE\alpha}}{\dpa\nabla_j \rho_\sigma}$ is given by:

\begin{equation}
    \begin{aligned}
     \label{eq:dpagradrho}
        \frac{\dpa  \varepsilon_{x,\sigma}^{PBE\alpha}}{\dpa\nabla_j \rho_\sigma} &= \frac{\dpa}{\dpa\nabla_j \rho_\sigma}\left(\rho_\sigma\varepsilon_{x,\sigma} F_{x,\sigma}^{PBE_\alpha}\right)\\
        &= \rho_\sigma \cdot \varepsilon_{x,\sigma} \cdot \frac{\dpa}{\dpa \nabla_j \rho_\sigma} F_{x,\sigma}^{PBE_\alpha}\\
        &= \rho_\sigma \cdot \varepsilon_{x,\sigma} \cdot \frac{\dpa}{s_\sigma} F_{x,\sigma}^{PBE_\alpha}\cdot \frac{\dpa s_\sigma}{\dpa \nabla_j \rho_\sigma}\,,
    \end{aligned}
\end{equation}

where $\frac{\dpa}{s_\sigma} F_{x,\sigma}^{PBE_\alpha}$ is given by:
\begin{equation}
    \begin{aligned}
     \label{eq:dpareducedgrad}
        \frac{\dpa}{s_\sigma} F_{x,\sigma}^{PBE_\alpha} &= (2\mu s_\sigma)\left(1 + \frac{\mu s_\sigma^{2}}{\kappa\alpha}\right)^{-1-\alpha}
    \end{aligned}
\end{equation}

and $\frac{\dpa s_\sigma}{\dpa \nabla_j \rho_\sigma}$ is given by:
\begin{equation}
    \begin{aligned}
     \label{eq:dpasgradrho}
        \frac{\dpa s_\sigma}{\dpa \nabla_j \rho_\sigma} &= \frac{(\nabla_j \rho_\sigma)}{2(6\pi^2)^{1/3}\rho_\sigma^{4/3} |\grad \rho_{\sigma}|}\,.
    \end{aligned}
\end{equation}

Therefore, combining Eq. \ref{eq:dparho}-\ref{eq:dpasgradrho} leads to the following expression for exchange potential of PBE$\alpha$ functional:

\begin{equation}
\label{eq:finalPBEalphaeq}
    \begin{aligned}
        V_{x,\sigma}^{PBE\alpha}  &= \varepsilon_{x,\sigma}\left(\frac{4}{3}F_{x\sigma} +\mu\left(1+\frac{\mu s_\sigma^2}{\kappa\alpha}\right)^{-\alpha - 1}\left(\frac{-8s_\sigma^2}{3}\right) \right)\\
        &- \sum_{j=1}^{3}  \left[\nabla_j\left(\frac{ (\nabla_j \rho_\sigma) \rho_\sigma\varepsilon_{x\sigma} \mu s_\sigma}{ (6\pi^2)^{1/3} \rho_{\sigma}^{4/3}  |\grad \rho_{\sigma}|} \left(1+\frac{\mu s_\sigma^2}{\kappa\alpha}\right)^{-1 - \alpha}\right)\right]\,.
    \end{aligned}
\end{equation}

\section{Separation of metallic and molecular environments for monopole and dipole features at different cut-off radius}

\begin{figure*}
    \centering
    \includegraphics[keepaspectratio=true,scale=0.425,clip=True]{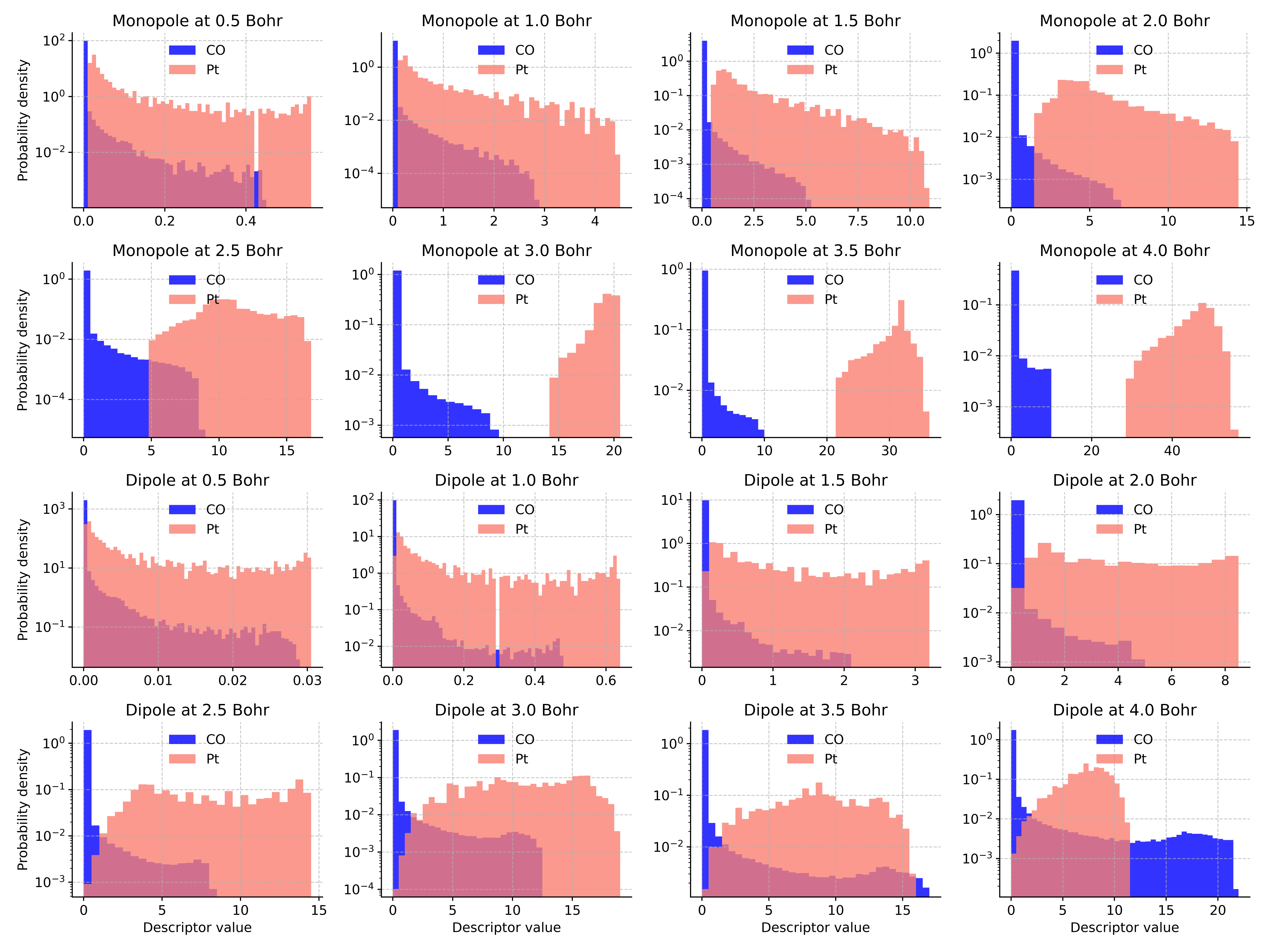}%wid\thetah=1.\linewid\thetah
    \caption{Histogram of monopole and dipole features illustrating separation between metallic (Pt) and molecular environments (CO) at different radial cut-offs ranging from 0.5-4 Bohr.}\label{fig:separation_histogram}
\end{figure*}

\section{Computational details for DFT calculations}

\begin{figure*}
    \centering
    \includegraphics[keepaspectratio=true,scale=0.55,clip=True]{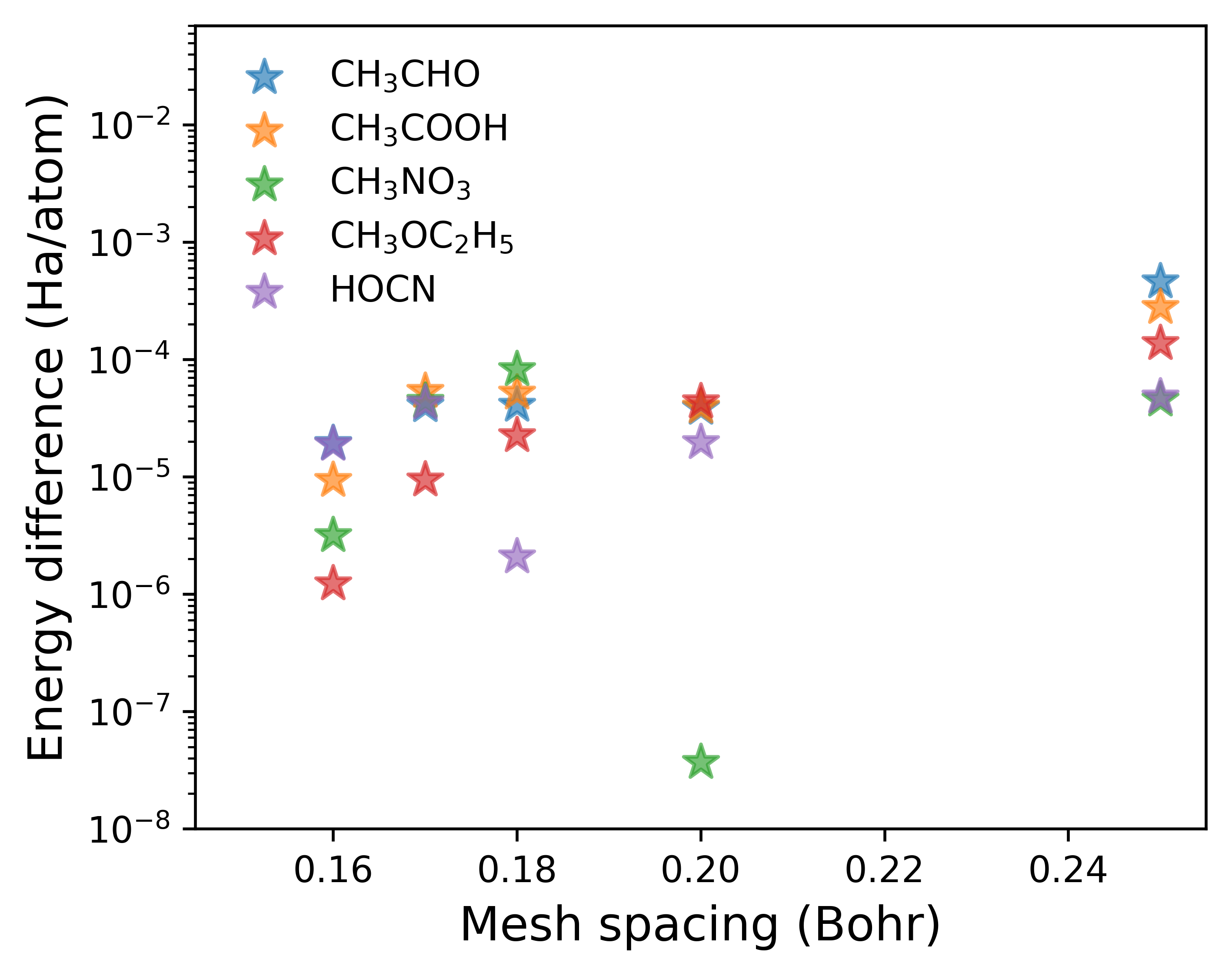}%wid\thetah=1.\linewid\thetah
    \caption{Convergence of total energy per atom for PBEq15 functional with respect to grid spacing. The energy difference/atom is calculated with respect to total energy at mesh spacing = 0.15 Bohr and is always below 0.001 Ha/atom.}\label{fig:mesh_convergence}
\end{figure*}

All DFT calculations are done using finite-difference DFT code, SPARC. We use periodic boundary conditions for all systems. For molecules, a vacuum of 5 \AA ~is used in each coordinate direction. For slabs, a vacuum of 10 \AA ~is used in z-coordinate direction. For Brillouin zone integration, we use 16 $\times$ 16 $\times$ 16 for lattice constant, 12 $\times$ 12 $\times$ 12 for cohesive energy and 4 $\times$ 4 $\times$ 1 Monkhorst-Pack grid for adsorption energy calculations.~\cite{monkhorst1976special}. The mesh spacing for all functionals including PBEq15 is chosen to provide a numerical accuracy of at least 0.001 Ha/atom and convergence testing is shown in Fig. \ref{fig:mesh_convergence}. The examples for input files for molecules, metals and chemisorption systems can be found in the GitHub repository. 

\section{Evaluation of self-consistent energies with and without convolution term}

\begin{figure*}[h]
    \centering
    \includegraphics[keepaspectratio=true,scale=0.5,clip=True]{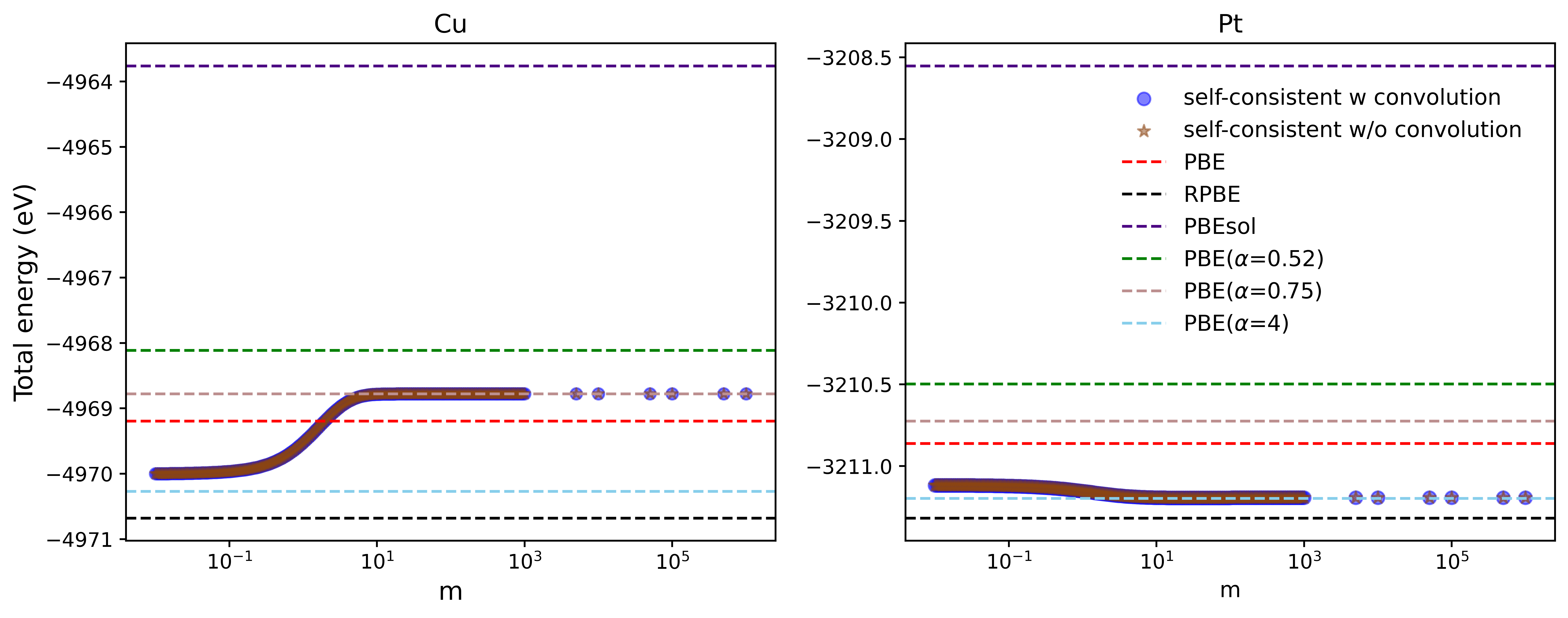}%wid\thetah=1.\linewid\thetah
    \caption{Effect of increasing m while keeping n fixed is illustrated for Cu and Pt where m is varied from $\mathrm{10^{-2}}$ to $\mathrm{10^6}$ and n = 20. The blue and brown scatter points are self-consistent energies with and without the convolution term, and are visually indistinguishable.}\label{fig:energy_sigmoid}
\end{figure*}

\section{Tabulated data for mean absolute error (MAE) and different properties of molecules and metals for different XC functionals}

\begin{table}
	\begin{center}
	\caption{MAE in formation energy w.r.t CCSD(T) for molecules, cohesive energy and lattice constants w.r.t experiment for metals.}\label{tab1}
		\begin{tabular}{lccc}	
\toprule		
Functional & Formation energy [eV] & Cohesive Energy [eV/atom] & Lattice constant [\AA]\\
\midrule
PBE & 0.169 & 0.274 & 0.043\\
RPBE & 0.161 &  0.598  & 0.093\\
PBEsol  & 0.227   &  0.566  &  0.026\\
PBE($\alpha=0.52$0)  &  0.189  &  0.265  & 0.034\\
PBE($\alpha=0.75$)	&  0.176  &  0.374  & 0.037\\
PBE($\alpha=4$) & 0.160 &  0.488  & 0.076\\
PBEq15 & 0.158  & 0.813 & 0.062\\
\bottomrule	
    \end{tabular}
        \end{center}
\end{table}

\begin{table*}
	\begin{center}
	\caption{Cohesive energy (eV/atom) for metals from different functionals.}\label{tab2}
		\begin{tabular}{lcccccccc}	
\toprule		
Metal & PBE & RPBE & PBEsol & PBE($\alpha=0.52$) & PBE($\alpha=0.75$) & PBE($\alpha=4$) & PBEq15 & Experimental\\
\midrule

Li & 1.588 & 1.522 & 1.6576 & 1.6338 & 1.6078 & 1.5389 & 1.5386 & 1.6633 \\
Na & 1.0453 & 0.9897 & 1.1127 & 1.0838 & 1.0597 & 1.0022 & 1.0022 & 1.1283 \\
K & 0.8013 & 0.7334 & 0.8658 & 0.8432 & 0.818 & 0.7524 & 0.7528 & 0.9428 \\
Rb & 0.6882 & 0.6149 & 0.7534 & 0.7332 & 0.7057 & 0.636 & ,0.6361 & 0.8574 \\
Ca & 1.8864 & 1.7028 & 2.0777 & 1.9995 & 1.9317 & 1.7577 & 1.7577 & 1.8623 \\
Sr & 1.5774 & 1.3836 & 1.7785 & 1.6971 & 1.6252 & 1.4414 & 1.4414 & 1.7343 \\
Ba & 1.7921 & 2.2281 & 2.6866 & 1.9394 & 1.8514 & 1.6225 & 1.6225 & 1.9107 \\
V & 5.0356 & 4.5344 & 5.6837 & 5.334 & 5.1561 & 4.6849 & 3.99 & 5.3468 \\
Nb & 6.8906 & 6.3207 & 7.5346 & 7.2303 & 7.0271 & 6.4927 & 5.9422 & 7.5967 \\
Ta & 8.1657 & 7.6192 & 8.8348 & 8.506 & 8.3022 & 7.781 & 7.1195 & 8.1233 \\
Mo & 6.2388 & 5.6073 & 11.027 & 6.6194 & 6.3898 & 5.7976 & 5.4143 & 6.8636 \\
W & 8.2007 & 7.5058 & 8.9388 & 8.6099 & 8.3642 & 7.717 & 7.1199 & 8.9388 \\
Rh & 5.8521 & 5.1812 & 6.702 & 6.2694 & 6.0185 & 5.3798 & 4.7508 & 5.7965 \\
Ir & 7.3738 & 6.6729 & 8.3542 & 7.8232 & 7.5513 & 6.8776 & 6.3291 & 6.9807 \\
Pd & 3.7508 & 3.1057 & 4.4747 & 4.1434 & 3.9088 & 3.2984 & 2.5335 & 3.9166 \\
Pt & 5.7415 & 5.0835 & 6.5816 & 6.155 & 5.9068 & 5.2776 & 5.5593 & 5.8633 \\
Cu & 3.5656 & 3.0234 & 4.1362 & 3.8849 & 3.694 & 3.1869 & 1.7235 & 3.5233 \\
Ag & 2.5321 & 1.9842 & 3.0794 & 2.8559 & 2.6642 & 2.1492 & 1.1921 & 2.9718 \\
Au & 3.0615 & 2.4498 & 3.7249 & 3.4338 & 3.2117 & 2.6318 & 1.8489 & 3.826 \\
Al & 3.4101 & 3.1634 & 3.7472 & 3.5661 & 3.4714 & 3.2365 & 3.2365 & 3.4315 \\
Pb & 2.8632 & 2.4271 & 3.2235 & 3.1108 & 2.9663 & 2.5604 & 1.7905 & 2.0402 \\
C & 7.6183 & 7.2316 & 8.144 & 7.8565 & 7.7117 & 7.3451 & 7.3451 & 7.5862 \\
Si & 4.4389 &  4.1592 & 4.7803 & 4.6113 & 4.508 & 4.2424 & 4.2424 & 4.6925 \\
\bottomrule	

\end{tabular}
\end{center}
\end{table*}

\begin{table*}
	\begin{center}
	\caption{Lattice constant (\AA) for metals from different functionals.}\label{tab2}
		\begin{tabular}{lcccccccc}	
\toprule		
Metal & PBE & RPBE & PBEsol & PBE($\alpha=0.52$) & PBE($\alpha=0.75$) & PBE($\alpha=4$) & PBEq15 & Experimental\textsuperscript{\cite{BEEF-vdW}}\\
\midrule
Li & 3.4457 & 3.4806 & 3.4517 & 3.4255 & 3.4378 & 3.4706 & 3.4706 & 3.451 \\
Na & 4.218 & 4.2999 & 4.2026 & 4.1705 & 4.1978 & 4.2771 & 4.2771 & 4.209 \\
K & 5.286 & 5.4409 & 5.2188 & 5.2102 & 5.2539 & 5.3895 & 5.3895 & 5.212 \\
Rb & 5.6708 & 5.8594 & 5.5725 & 5.576 & 5.6313 & 5.7972 & 5.7972 & 5.577 \\
Ca & 5.521 & 5.6154 & 5.4537 & 5.4706 & 5.4997 & 5.5852 & 5.5852 & 5.556 \\
Sr & 6.0315 & 6.1513 & 5.9247 & 5.9627 & 6.0035 & 6.1143 & 6.1143 & 6.04 \\
Ba & 5.02 & 5.1497 & 4.8787 & 4.943 & 4.9889 & 5.1098 & 5.1098 & 5.002 \\
V & 2.9993 & 3.0224 & 2.9664 & 2.9849 & 2.9929 & 3.0145 & 2.9929 & 3.024 \\
Nb & 3.311 & 3.3302 & 3.2728 & 3.2975 & 3.3055 & 3.3249 & 3.3055 & 3.294 \\
Ta & 3.3197 & 3.3385 & 3.2851 & 3.307 & 3.3147 & 3.3331 & 3.3148 & 3.299 \\
Mo & 3.1623 & 3.1762 & 3.1321 & 3.1526 & 3.1585 & 3.1723 & 3.1584 & 3.141 \\
W & 3.1839 & 3.1961 & 3.157 & 3.1753 & 3.1806 & 3.1927 & 3.1806 & 3.16 \\
Rh & 3.8311 & 3.8553 & 3.7834 & 3.8152 & 3.8248 & 3.8482 & 3.8248 & 3.793 \\
Ir & 3.8692 & 3.8859 & 3.8284 & 3.8574 & 3.8648 & 3.8814 & 3.8648 & 3.831 \\
Pd & 3.9473 & 3.9887 & 3.882 & 3.9218 & 3.9368 & 3.9767 & 3.9368 & 3.876 \\
Pt & 3.9698 & 3.9942 & 3.9169 & 3.9529 & 3.9634 & 3.9871 & 3.9634 & 3.913 \\
Cu & 3.6143 & 3.6867 & 3.5655 & 3.5931 & 3.6056 & 3.6594 & 3.6056 & 3.596 \\
Ag & 4.1458 & 4.2251 & 4.0545 & 4.1041 & 4.1287 & 4.2014 & 4.1289 & 4.062 \\
Au & 4.1579 & 4.2025 & 4.0821 & 4.1294 & 4.1465 & 4.1897 & 4.1466 & 4.062 \\
Al & 4.0387 & 4.0408 & 4.0246 & 4.0369 & 4.038 & 4.0403 & 4.0403 & 4.019 \\
Pb & 5.0402 & 5.1257 & 4.9287 & 4.9885 & 5.0194 & 5.1 & 5.0918 & 4.912 \\
C & 3.5742 & 3.5808 & 3.5582 & 3.5699 & 3.5725 & 3.5792 & 3.5778 & 3.544 \\
Si & 5.4688 & 5.4813 & 5.4456 & 5.4604 & 5.4656 & 5.4776 & 5.4776 & 5.415 \\
\bottomrule	
\end{tabular}
\end{center}
\end{table*}

\section{Adsorption energy for CO on Pt(111)}
\begin{figure*}
    \centering
    \includegraphics[keepaspectratio=true,scale=0.8,clip=True]{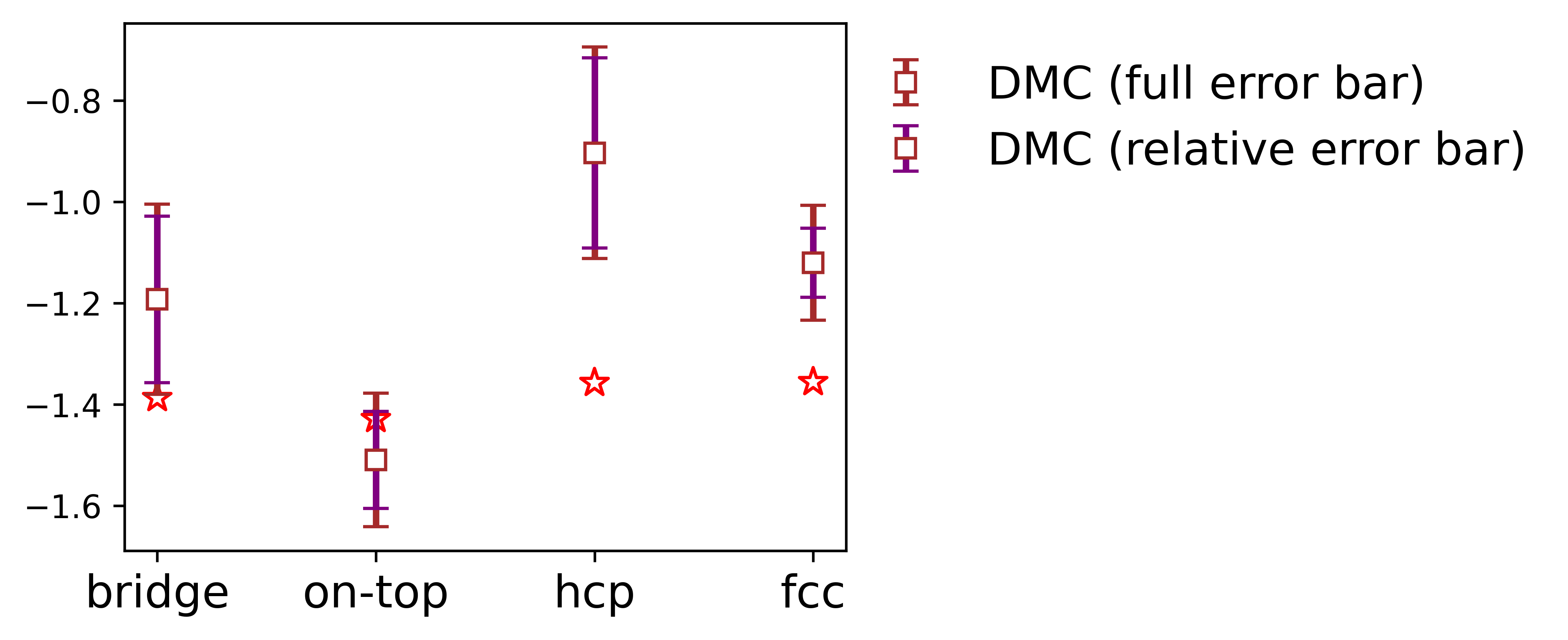}%wid\thetah=1.\linewid\thetah
    \caption{Adsorption energy on four sites: bridge, on-top, hcp and fcc. Results from PBEq15 functional and DMC with full and relative error bars are included.}\label{fig:adsorption_energy_err}
\end{figure*}

In Fig. \ref{fig:adsorption_energy_err}, we show the DMC adsorption energies with two sets of error bars. The larger (maroon) error bars correspond to the overall statistical error of the binding energy calculations. However, we note here that the DMC energies of the pure Pt(111) slab and the isolated CO molecule contribute a fixed additive error to each of the four adsorption energy calculations. Since this fixed error does affect the DMC energy of the slab/adsorbate systems themselves, it can be ignored when one is only interested in establishing the relative preference of the adsorption sites and not the absolute error of the binding energy. Ignoring this fixed additive error results in a set of smaller (purple) error bars in Fig. \ref{fig:adsorption_energy_err}. This unambiguously identifies the on-top site as the preferred adsorption site in the CO/Pt(111) system at the DMC level.

% table for adsorption energy
\newpage
\setlength{\bibsep}{0.0cm}
\bibliographystyle{Wiley-chemistry}
\bibliography{example_refs}